\documentclass[pdflatex,sn-mathphys-num]{sn-jnl}

\usepackage{graphicx}%
\usepackage{multirow}%
\usepackage{amsmath,amssymb,amsfonts}%
\usepackage{amsthm}%
\usepackage{mathrsfs}%
\usepackage[title]{appendix}%
\usepackage{xcolor}%
\usepackage{textcomp}%
\usepackage{manyfoot}%
\usepackage{booktabs}%
\usepackage{algorithm}%
\usepackage{algorithmicx}%
\usepackage{algpseudocode}%
\usepackage{listings}%
\usepackage{setspace}

\usepackage{listings}
\usepackage{longtable}
\usepackage{geometry}
\usepackage{pgfplots}
\pgfplotsset{compat=newest}
\usepackage{parskip}
\usepackage{verbatim}
\usepackage{pdflscape}
\usepackage{tabularx}
\usepackage{ltablex}
\usepackage{hyperref}
\usepackage{array}
\usepackage{enumitem}
\usepackage{forest}
\usepackage{tikz}
\usepackage{subcaption}
\usepackage{threeparttable}

\theoremstyle{thmstyleone}%
\theoremstyle{thmstyletwo}%

\theoremstyle{thmstylethree}%

\begin{document}

\title[Article Title]{Connecting Distributed Ledgers: Surveying Novel Interoperability Solutions in On-chain Finance}

\author*[1,2]{\fnm{Hasret Ozan} \sur{Sevim}}\email{hasretozan.sevim@unicatt.it}

\affil[1]{University of Camerino (Camerino, Italy)}

\affil[2]{Catholic University of Sacred Heart (Milan, Italy)}

\abstract{This paper emphasizes the critical role of interoperability in enabling efficient and secure communication for the fragmented distributed ledger ecosystem, particularly within on-chain finance. The purpose of this study is to streamline and accelerate empirical research on the intersection of cross-chain interoperability solutions and their impact within on-chain finance. The analysis examines the relationship between financial use and interoperability while comparing the properties of novel cross-chain interoperability protocols (LayerZero, Wormhole, Connext, Chainlink Cross-Chain Interoperability Protocol, Circle Cross-chain Transfer Protocol, Hop Protocol, Across, Polkadot, and Cosmos), focusing on their design, mechanisms, consensus, and limitations. To encourage further empirical study, the paper proposes a set of network metrics and sample statistical models and provides a framework for evaluating the performance and financial implications of interoperability solutions.}


\maketitle

\section{Overview}\label{sec1}

The landscape of cross-chain interoperability solutions has gained significant attention as the distributed ledger network ecosystem keeps growing and becomes more fragmented. The need for interoperability arises from the idea of benefiting from the strengths of multiple blockchains simultaneously, creating more integrated, efficient, and versatile systems. Ideal interoperability solutions aim to overcome issues like liquidity fragmentation across different ledgers, limited accessibility to diverse assets, operational inefficiencies, increased counterparty risks, lower financial adoption and innovation across the distributed ledger technologies (DLT) ecosystem, and poor user experience. The main goal of cross-chain interoperability is to enhance efficiency, scalability, and versatility in asset and data transfers within the blockchain ecosystem \cite{Harris2023}. The literature is rich in examining, defining, and sorting different cross-chain interoperability solutions\cite{Wang2023}. There are studies on the design patterns and solutions (Hyperledger Cactus, Cosmos, Polkadot, Interchain, and Interledger). However, fewer studies exist on measuring the performance and efficiency of those cross-chain interoperability solutions and their relationship with on-chain finance protocols (such as how cross-chain interoperability solutions affect the performance and financial efficiency of automated market makers and automated lending protocols). Empirical studies are mostly limited by cross-chain arbitrage. The literature lacks comparative studies on the currently widely adopted multi-chain interoperability solutions (LayerZero, Chainlink’s CCIP, Circle’s CCTP, Wormhole, Connext, Hop Protocol, and Across) and empirical studies on the financial aspects of these solutions.

This paper is written to detect the developments in the DLT interoperability solutions, explore the capabilities and technical mechanisms of these solutions, and examine how the financial output of those interoperability solutions can be empirically analyzed. The aim of this paper is not only to provide a technical comparison of the novel interoperability solutions but, more importantly, to attract the attention of the financial experts and academics to analyze the financial outputs and aspects of these solutions. In other words, this study provides knowledge, mechanisms and a comparison of different cross-chain interoperability solutions, an example of an empirical method to address the direction and necessity of future research in this field. The purpose of this paper is to be a ground for further research for financial analysis of multi-chain interoperability solutions and their impacts within the on-chain finance ecosystem and protocols. In other words, this paper fills a gap by being a pipeline between current cross-chain interoperability solutions, current studies, and future empirical research. The scope of this paper does not cover native bridges that are designed for the communication of two blockchains (a layer-1 and a layer-2 blockchain), such as Arbitrum Bridge, but multi-blockchain and inter-blockchain interoperability solutions.

Section \ref{sec2} of the paper takes a look at the concepts of DLTs and interoperability. Section \ref{sec3} examines the financial use cases provided by cross-chain interoperability. Section \ref{sec4} scans the existing literature. Section \ref{sec5} sorts the promising and popular multi-chain interoperability solutions and examines their features, designs and mechanisms. These solutions are LayerZero, Wormhole, Connext, Circle's Cross-chain Transfer Protocol (CCTP), Chainlink's Cross-chain Interoperability Protocol (CCIP), Hop Protocol, Across, Polkadot, and Cosmos. Section \ref{sec6} compares these solutions from many aspects: architecture, security model, modularity, trust setup, chain agnosticism, and limitations. Section \ref{sec7} includes basic statistical charts of LayerZero and tables of Wormhole to demonstrate the rising interest in the usage of cross-chain solutions. Section \ref{sec8} introduces statistical methods to exemplify an analysis of the financial data produced by those cross-chain solutions. Section \ref{sec9} concludes the paper and shows a potential path for future research. Across the paper, the term 'distributed ledger technologies' or 'distributed ledger networks' might be used instead of blockchain in case some different systems are also integrated for interoperable communication.

\section{Key Concepts}\label{sec2}

Understanding cross-chain interoperability solutions requires knowledge of fundamental concepts of distributed ledger technologies, especially blockchain. In this paper, these fundamental concepts are not referred to in detail, instead shortly. 

In a broad sense, blockchain can be explained as an immutable, considerably more decentralized, trusted, and distributed ledger based on peer-to-peer networks. Essentially, blockchain is a data structure that functions to record transactions interdependently generated within a network where the distributed ledger is constantly synchronized between peers. Blockchain can be categorized into distinct categories according to how blockchain organizes its participants in different application scenarios. Permissionless blockchains allow anyone to run the blockchain node with full transparency, while permissioned blockchains incorporate only pre-approved nodes \cite{Punathumkandi2021}. Blockchains can provide scripting environments to enable users to deploy software programs known as 'smart contracts' that can be triggered by a transaction of a user \cite{Xu2019}. A smart contract is a self-executing digital contract stored on a blockchain, automatically enforcing predefined terms and conditions without the need for intermediaries. Smart contracts disintermediate traditional intermediaries. Cross-blockchain smart contracts target general blockchain interoperability differently from cross-blockchain token transferring mechanisms.

On-chain smart contracts require off-chain data to enable further use cases beyond the capabilities of isolated blockchains. For instance, financial smart contracts may require the price data of some commodities for their business application. In that case, those smart contracts import external data from off-chain data providers 'oracles' \cite{Ellis2017}. Oracle networks may have decentralized or centralized decision-making architectures. Although oracles are not inherently cross-chain solutions, they can play a significant role in interoperability frameworks by providing necessary external data that can facilitate interactions between chains and validate transactions across chains by providing relevant external proofs. So oracle service and cross-chain interoperability can be considered complementary elements in some cases.

At this point, it is requisite to explain what on-chain finance (also known as 'decentralized finance') mean. On-chain finance is the sum of protocols that provide the usage of banking services and other miscellaneous financial services without any bank or intermediary \cite{Schar2021}.

Innovations in distributed ledger technologies, such as atomic swaps, relay chains, and rollup solutions, have provided new mechanisms to facilitate cross-chain interactions, contributing to the rise of interoperability solutions. Despite the significant progress, cross-chain interoperability solutions face challenges that could hinder their widespread adoption and impact. Interoperability increases the vulnerability of blockchain systems that can be exploited across chains with more catastrophic results \cite{Duan2023}. Ensuring the security of cross-chain transactions remains a significant challenge. Developing protocols that try to enable seamless interaction between different blockchains is technically complex. These solutions must be scalable to handle growing transaction volumes without compromising performance or security. The lack of standards and governance models for cross-chain interoperability complicates integration efforts.

Interoperability is the ability of two or more systems to exchange information despite their differences in language, interfaces, and execution platform \cite{Wegner1996}. Originally, blockchains are isolated information silos. Thus, they require additional technical capabilities to accept data from another distributed ledger system with a trustless consensus. Because having interoperability capability with trusted third parties may create conflict with the desired decentralization level.

Thus, cross-chain interoperability can be defined as a general term for the ability to transfer data and tokens (including tokenized assets) between different blockchain networks by maximizing decentralized ways and mechanisms. Interoperability can be provided in different ways, like designing blockchains with similar mechanisms from the beginning and connecting them with a protocol ('interblockchain' approach) or orchestrating transactions via gateway applications on multiple blockchains. Multi-blockchain interoperability solutions are also known as 'multi-chain bridges'.

Several interoperability solutions exist to execute cross-chain transactions \cite{Belchior2022} \cite{Llambias2023}. Notary scheme is a solution in which a trusted third party monitors the source ledger network to submit transactions to a destination network. Another solution is sidechain that are widely adopted from cross-chain communication to layer-2 networks. Sidechains connect blockchains with two-way peg mechanisms that enable the transfer of assets between the sidechain and the mainchain. In reality, sidechains exist outside of a blockchain and they can have different consensus protocols or implementations. Similar to sidechains, roll-up blockchains also require cross-chain blockchain communication and ultimately settle on a main (layer-1) network as another layer-2 network solution. Relays are smart contracts specifically deployed on both blockchains to verify the received transactions. Gateways are middleware protocols used by relays to use proofs of verification. Relay chains can be used as central hubs that facilitate communication and transactions between different blockchains (parachains) connected to the network. A big-scale solution is to create a blockchain of blockchains. The role of the mainchain is sophisticated to provide the execution of cross-chain transactions between several blockchains. Blockchain of blockchains is a generalization of the sidechain/relay solution, comprising a main chain and multiple sidechains collaborating with each other. It is a trustless solution but at the cost of high complexity \cite{Llambias2023-2}.

\begin{table}[h]
\caption{A typical cross-chain interoperability protocol has this structure.}\label{tab1}%
\begin{tabular}{@{}ll@{}}
\toprule
\textbf{Layer} & \textbf{Function} \\
\midrule
Transport & Read the root or hash of data on the origin chain and post to the destination. \\
Verification & Ensure that the posted data is correct. \\
Execution & Generate Merkle roots on origin. \\
 & Prove against thee root and execute the target function. \\
Application & Handle specific use cases such as token transfers and decentralized governance. \\
\botrule
\end{tabular}
\end{table}

Modularity has gained importance in the domain of DLT interoperability. Modularity refers to a composable and flexible design or architecture in different layers of interoperability solutions. Applications can customize intent layers, relayer networks, settlement layers or oracle networks. However, this customizable nature creates more complexities in comparison with monolithic systems.

\begin{table}[h]
    \centering
    \caption{The map of cross-chain interoperability components are presented.}\label{tab2}
    \label{tab:cross-chain-interoperability}
    \begin{tabular}{|l|l|p{7cm}|}
        \hline
        \textbf{Component Type} & \textbf{Components} & \textbf{Details} \\ \hline
        \multirow{4}{*}{Infrastructure patterns} 
        & Blockchain API Gateway & \multirow{4}{6cm}{Enhances efficiency of cross-chain application patterns. Enables data transfer to data migration patterns and cross-chain application patterns.}\\
        & Relayer & \\ 
        & Aggregator & \\
        & & \\ \hline
        Cross-chain application patterns 
        & Temporal transfer & Assets are temporarily moved between blockchains. \\ \hline
        Data migration patterns 
        & Permanent transfer & Assets are permanently moved from one blockchain to another. \\ \hline
        \multirow{2}{*}{Security patterns} 
        & Light client & Enhances cross-chain verification to all other components. \\
        & & \\ \hline
    \end{tabular}
\centering
\footnotetext{Source: Llambías et al. (2024) \cite{Llambias2024}}
\end{table}

The transactions of exchanges on DLTs require an atomic swapping process, which can guarantee the integrity of different execution processes. Different parties can trade their assets from different blockchains with each other thanks to atomic swaps. Both parties should have an address on the other blockchain, and the trades must happen simultaneously on both blockchains. Both transfers must be guaranteed to happen or neither of them happens. This property is called 'atomic', as the swap process is made as a whole. The atomic swap can be adopted into multiple blockchain scenarios, which is referred to as an atomic cross-chain swap \cite{Robinson2021}. An atomic cross-chain swapping process is a distributed coordination task that enables the exchange of assets across multiple blockchains atomically \cite{Wang2023}. The whole swapping process is executed and automated with the help of smart contracts. Different methods exist to implement atomic cross-chain transactions. One way is the use of Hash Time-Locked Contracts (HTLC) \cite{Bishnoi2022}. HTLCs are a type of time-bound smart contract that can be used in cross-chain transactions, especially for atomic swaps. This type of contract ensures that a transaction between parties will be completed only if all conditions are met within a specific time frame. Otherwise, the transaction is nullified and assets are returned to their original owners.

Cross-chain communication is one of the most important design considerations in the current DLT-based systems. Because each blockchain system operates as an isolated zone of information, where it is difficult to obtain external data. Each blockchain executes transactions on its own. But cross-chain bridges are specialized protocols to provide a transfer service of assets and information between different blockchain networks. Bridges can be classified differently depending on their features, including trusted bridges and trustless bridges. Trusted bridges rely on trusted entities or consortiums to validate and facilitate transactions between blockchains. It is easier to implement trusted bridges as those bridges do not require a sensitivity of high decentralization. However decentralized bridges are designed for reduced trust and centralization. They can be more complex and must ensure security across disparate systems\cite{Wang2023}. At this point, the risks and vulnerabilities of cross-chain bridges should be mentioned. An attack or outage in one chain can affect the ongoing cross-chain transactions and the assets in the bridge contracts. Cross-chain bridges may have centralized components, such as custodial wallets or oracle services, which may introduce security risks. Bridge contracts can be susceptible to exploits and hacks, which may result in the loss of user funds. The bridges may have higher fees compared to DEXs, as they may require additional infrastructure and services to operate effectively\cite{Harris2023}.

Wrapped tokens represent native assets on a different blockchain non-natively. They are created to enable cross-chain compatibility by locking the original asset on its native chain and issuing a corresponding token on another chain \cite{Caldarelli2022}. Burn/lock and creation of assets are mechanisms to enable cross-chain asset transfer, exchange, and data sharing with atomic tasks between blockchains to ensure the prevention of double-spending.

It can be supposed that an ideal protocol, which is designed to enable transactions between blockchains, should have the following properties \cite{Gao2022}:
\begin{itemize}
    \item Permissionless: Every node of a distributed ledger are allowed to join the communication network.
    \item Safety: All state updates for a single transaction are either successful or fail on all involved blockchains.
    \item Liveness: All valid transactions will eventually succeed on all blockchains.
    \item Linearizability: All parallel successful transactions is always linearizable so that the consequent state of a group.
\end{itemize}

It might be beneficial to underline the difference between trusted and trustless systems as different systems are used in interoperability solutions as well. Trustless systems minimize the need for trust among participants by relying on technology and mathematics to secure transactions, whereas trusted systems depend on the credibility and reliability of central authorities \cite{Harz2018}. In a trustless system, transactions and interactions are conducted with a considerably low level of trust between participants. This feature typically can be achieved by the technique of deterministic execution. Instead of relying on a central authority or intermediary to validate transactions, trustless systems can rely on cryptographic algorithms and consensus mechanisms as much as possible to ensure the integrity and security of transactions. Consensus mechanisms allow a distributed network of participants (nodes) to agree on the validity of transactions and the state of the ledger without needing to trust each other. Trustless systems may also promote transparency and inclusivity, as anyone with the necessary resources can participate in the network as long as the network activity does not require a certain level of trust and confidentiality. However, participants in a trusted system must have confidence in these central entities to act honestly and competently. In trusted DLT systems, a predefined set of validators (which could be financial institutions, corporations, or consortia) is responsible for validating transactions and maintaining the ledger \cite{Bakos2021}. Trusted systems can offer greater scalability and efficiency compared to trustless systems, as the consensus process does not typically require complex consensus mechanisms. Trusted systems provide more control over the network, especially for compliance concerns like regulation and privacy.

Cross-chain interoperability can be used for different use-cases. Since distributed ledger technologies are mainly used for financial cases, the current usage of cross-chain interoperability solutions is mostly related to financial activities.

\section{Relationship Between Interoperability, Business Applications and On-chain Finance}\label{sec3}

For on-chain finance applications, interoperability could significantly enhance liquidity by pooling resources across multiple blockchains. It could also provide users and businesses with access to a broader range of assets, a combination of different smart contract use-cases, and services, potentially increasing participation in the digital economy. At this point, cross-chain interoperability becomes more than an infrastructural topic with different financial use-cases and business models.

In recent years, organizations started building consortium blockchains to improve their business processes. They consider blockchain as a suitable technology that guarantees immutability, confidentiality, and availability of a common data source to every party in the consortium. Although there have been significant developments improving blockchain platforms, they still operate independently from each other and cannot be integrated easily, leading to information silos \cite{Llambias2023-2}. Permissioned blockchains like Hyperledger can be used by enterprises for their internal operations due to their privacy controls and scalability. These could interoperate with public blockchains to provide transparency and traceability to end users for certain aspects of the provided services and operations, such as the origin of funds/raw materials, leveraging the controlled environment of permissioned DLTs and the transparency of public blockchains. Financial applications of cross-ledger solutions will also help feed financial needs with several liquidity sources. Recently, roll-up blockchains (a different type of 'layer-2' blockchains) have also been adopted by organizations and institutions to operate with permissionless DLTs better.

\begin{table}[h]
\caption{Cross-chain communication (CCC) enables a variety of financial use cases.}\label{tab:cross-chain-use-cases}
\begin{tabular}{@{}p{0.3\linewidth}p{0.6\linewidth}@{}}
\toprule
\textbf{Use Case} & \textbf{Function} \\
\midrule
Cross-chain exchange & Developers can build an exchange that allows deposits from any connected chain and withdrawals from another connected chain. This type of use case increases accessible liquidity and reduces liquidity fragmentation across different distributed ledger networks (aggregated liquidity). \\
\midrule
Cross-chain lending & CCC enables users to lend and borrow a wide range of crypto assets across multiple on-chain finance platforms running on independent ledger networks. \\
\midrule
Low-cost transaction computation & CCC can help offload the computation of transaction data on cost-optimized chains. \\
\midrule
Optimizing cross-chain yield & Users can move collateral to new on-chain finance protocols to maximize yield across chains. \\
\midrule
Creating new kinds of business models and applications & CCC enables users to take advantage of network effects on certain chains while harnessing the compute and storage capabilities of other chains. \\
\midrule
Cross-chain governance & A group of different consortiums on different networks can vote on a combined proposal if they would like to communicate votes cast on a common chain. \\
\botrule
\end{tabular}
\end{table}

\begin{figure}[htbp]
\centering
\includegraphics[width=\linewidth]{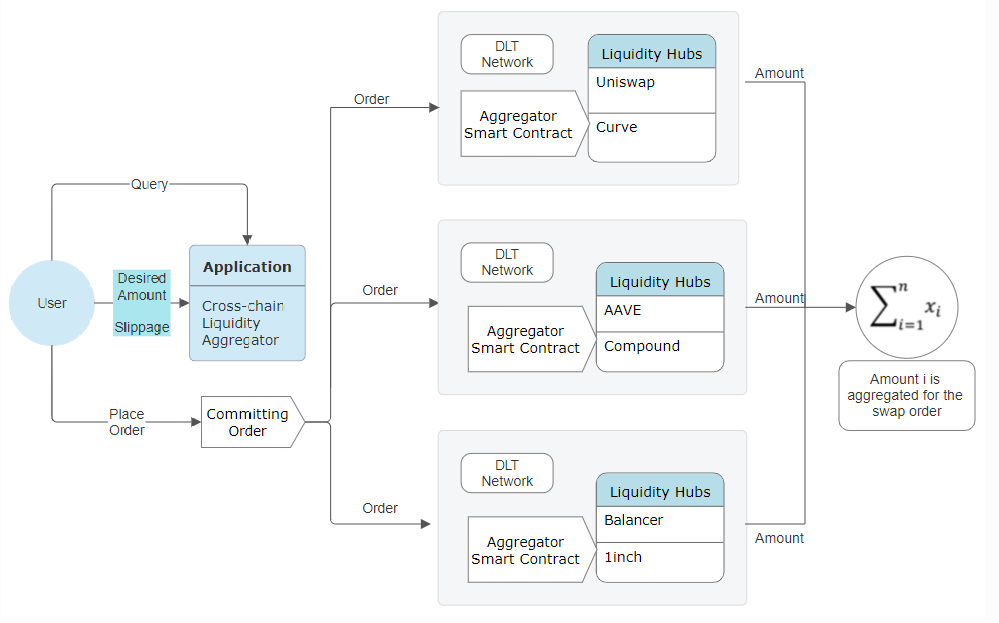}
\caption{The diagram illustrates a cross-ledger liquidity aggregation from different on-chain finance liquidity hubs at the same time.}
\label{fig:diagram2}
\end{figure}

Interoperability could lead to a more inclusive and collaborative blockchain space with communication and transactions across diverse blockchain ecosystems by breaking down silos and leveraging the unique strengths of different DLTs. Consider a scenario where a blockchain, that is known for its high security but slower transaction speeds, like Ethereum, interoperates with a more efficient but less secure blockchain. Assets could be seamlessly transferred between blockchains, leveraging the strengths of each. For example, a digital asset or token created on Ethereum could be transferred to a blockchain with lower transaction fees for everyday transactions and then moved back to Ethereum when needed for specific smart contract functionalities.

The notion of a multi-chain future, including permissioned blockchains and roll-up blockchains potentially operated by banks, is not just speculative but seems an almost inevitable reality. However the notion of a cross-chain and/or multi-chain future will be possible when the chronic issues inherent in the concept of permissionless cross-chain interoperability solutions are solved. A hybrid approach, that utilizes both permissioned and permissionless elements where they are most beneficial, might be the most effective solution. But still operating in a less trusted environment can be particularly pertinent for parties from completely different jurisdictions, due to several factors inherent in cross-jurisdictional interactions. There are key reasons of why a less trusted or trustless environment, often facilitated by permissionless blockchain platforms, might be required or beneficial in such cases:

\begin{itemize}
    \item Reduced Counterparty Risk: In international dealings, the risk of fraud, default, or non-compliance with agreements can be removed or reduced by trustless distributed ledger systems.
    \item Lack of Common Regulatory Framework: A considerably more decentralized and trustless system can provide a neutral ground where transactions and agreements are enforced by code, reducing the reliance on any single jurisdiction's legal framework.
    \item Cost and Efficiency: A trustless interoperability protocol and a considerably more decentralized blockchain can enable direct peer-to-peer interactions without the need for trusted intermediaries, automating transactions, recording them in an immutable and unaltered ledger, leading to more efficient and cost-effective transactions.
    \item Anonymity and Privacy: Parties from different jurisdictions may require or prefer anonymity, or at least a degree of privacy in their transaction. Permissionless blockchains can offer mechanisms to protect user identities and transaction details while still ensuring the integrity and verifiability of transactions.
    \item Interoperability and global access.
\end{itemize}

However, each distributed ledger system adds layers of complexity to cross-chain interactions with its unique consensus mechanism, governance model, and technical infrastructure. This complexity could lead to bottlenecks, increased transaction costs, and potential points of failure, undermining the very efficiencies blockchain technology seeks to offer. Interoperability solutions require some level of trust in bridging protocols or intermediary chains, which could become prime targets for attacks. The more complex and interconnected the ecosystem, the larger the attack surface becomes. The potential for exploits, hacks, and fraud increases exponentially with each added layer of interoperability, especially when dealing with permissioned blockchains that may have differing security standards or vulnerabilities.

A very practical example of a cross-chain business implementation is Pike Finance\footnotemark, which was a leading example of a multi-blockchain asset lending and borrowing application using Wormhole multi-blockchain interoperability infrastructure. It was possible to borrow e-money token (also known as 'stablecoins') on a blockchain by providing different collateral assets in another blockchain. By looking at this example, it is possible to speculate on an alternative international money market, like borrowing a tokenized EUR in a layer-2 network of a European bank by providing tokenized USD collateral in a different layer-2 network of a financial intermediary.

\footnotetext{The technical details of the application are given on its website: https://docs.pike.finance/. Pike Finance stopped operation after experiencing a security breach on 26th April 2024.}

\section{Existing Studies}\label{sec4}

In the literature, many studies exist regarding the theoretical and technical framework of interoperability issues and solutions for DLTs. Especially, architectures like sidechains, notary schemes, relay chains, parallel chains (also known as 'parachains'), rollups and critical technical features like atomic swap and hash-time locked smart contracts are well-studied and developed. Financial utilities, such as asset transfers, are analyzed. However the implications have been understudied other than Cosmos, Polkadot, Interledger, and Interchain. More comparative and empirical approach is required as the interoperability developments have been exponentially accelerated between 2020 and 2024, and the new solutions are evolving into hybrid architectures and applications.

The following studies give significant insights to understand cross-chain interoperability, the classification of different interoperability mechanisms, the room for improvement, and potential methodologies to improve the current state-of-the-art.

Belchior, Vasconcelos, Guerreiro, and Correia (2021) conduct a literature review by classifying the studies into three distinct categories as crypto-asset-directed interoperability approaches, blockchain engines and blockchain connectors \cite{Belchior2022}. Then the paper proposes the Blockchain Interoperability Framework (BIF), a framework defining criteria to assess blockchain interoperability solutions. Llambías, González, and Ruggia (2024) propose a work of design pattern identification to reveal blockchain interoperability problems from the practice \cite{Llambias2023}. Six interoperability patterns are identified through the observation of 35 interoperability solutions. The outputs are evaluated with five semi-structured interviews with blockchain experts. Kotey et al. (2023) offer a systematic review on heterogeneous blockchain-to-blockchain communication, discussing the current state and future research directions \cite{Kotey2023}. Zilnieks and Erins (2023) investigate the role of cross-chain bridges in standardizing DLT within payment systems, which could be crucial in creating a more efficient financial landscape\cite{Zilnieks2023}. Zamyatin et al. (2021) present a systematic approach to cross-chain communication (CCC) by laying out foundational work for future CCC protocols, which is critical for interoperable financial smart contracts \cite{Zamyatin2021}. Mafike and Mawela (2023) conduct a literature review on requirements for interoperable blockchain systems, focusing on technical, semantic, legal, and organizational aspects \cite{Mafike2023}. McCorry et al. (2021) review validating bridges as a scaling solution for blockchains, discussing operational costs and security implications \cite{McCorry2021}. Lee, Murashkin, Derka, and Gorzny (2022) provide a review of cross-chain bridge hacks, identifying risks and security improvements for these systems \cite{Lee2022}. Ren et al. (2023) provide a comprehensive survey on blockchain interoperability by categorizing current solutions and addressing the performance of some approaches \cite{Ren2023}. Pillai et al. (2020) explore the trade-offs between security and performance in blockchain interoperability, which impacts the decentralization and integrity of blockchain systems \cite{Pillai2022}. Pfister, Kannengießer, and Sunyaev (2022) discuss the balance between technical and political decentralization in token economies and the role of cross-ledger interoperability\cite{Pfister2022}. Ming et al. (2024) explore the fusion protocol’s cross-chain and interoperation methods, focusing on relay-chain technology and universal digital wallet concepts \cite{Ming2024}. Harris (2023) examines challenges and opportunities for blockchain interoperability, exploring various cross-chain technologies and projects \cite{Harris2023}. Han et al. (2023) conduct a survey on cross-chain technologies, proposing a blockchain interoperability architecture to address security and effectiveness issues \cite{Han2023}. The following studies deep dive to understand cross-chain interoperability concepts, terms, and literature: Bokolo (2022) \cite{Bokolo2022-2}, Pillai et al. (2020) \cite{Pillai2020}, Meng et al. 2022 \cite{Meng2022}, Falazi et al. (2024) \cite{Falazi2024}, Williams (2020) \cite{Williams2020}, Talib et al. (2021) \cite{Talib2021}, Robinson (2021) \cite{Robinson2021-2}, Jiang et al. (2023) \cite{Jiang2023}, Bayraktar and Gören (2023) \cite{Bayraktar2023}, and Wang (2022) \cite{Wang2022}.

Mazor and Rottenstreich (2023), an empirical study of cross-chain arbitrage in on-chain finance platforms, analyze profitability and challenges in blockchain ecosystems \cite{Mazor2023}. This study is one of the rare empirical studies focusing on the cross-chain landscape.

The literature involves many studies, focusing on the technical parts and mechanisms of cross-chain interoperability, proposing or developing tools and protocols, such as Yin, Xu and Zhang (2023) \cite{Yin2023}, Sober et al. (2023) \cite{Sober2023}, Robinson and Ramesh (2021)\cite{Robinson2021}, Pourpouneh et al. (2023) \cite{Pourpouneh2023}, Monika and Bhatia (2024)) \cite{Monika2023}, Miyaji and Yamamoto (2024) \cite{Miyaji2024}, Mars et al. (2023) \cite{Mars2023}, Madhuri and Nagalakshmi (2023) \cite{Madhuri2023}, Lu et al. (2023) \cite{Lu2023}, Llambias et al. (2023) 
 \cite{Llambias2023}, Li, Wu, and Cui (2023) \cite{Li2023}, Lee et al. (2021) 
 \cite{Lee2021}, Jnr. et al. (2023) \cite{Bokolo2023}, Jiang et al. (2023) \cite{Jiang2023}, Hei et al. (2022) \cite{Hei2022}, Guo et al. (2024) \cite{Guo2024}, Darshan et al. (2023) \cite{Darshan2023}, Chen et al. (2024) \cite{Chen2024}, Wang and Nixon (2021) \cite{Wang2021}, Barbara and Schifanella (2023) \cite{Barbara2023}, Vishwakarma (2023) \cite{Vishwakarma2023}, Westerkamp and Eberhardt (2020) \cite{Westerkamp2020}, Gao et al. \cite{Gao2022}, Xie et al. (2022) \cite{Xie2022}, and Tsepeleva and Korkhov (2022) \cite{Tsepeleva2022}.

\section{Novel Solutions}\label{sec5}
In this section, five cross-chain messaging protocols, two cross-chain token transfer protocols, and two cross-chain communication-focused blockchains are analyzed. Their smart contract orchestrations are also demonstrated with figures.

\subsection{LayerZero}\label{subsec1}

LayerZero is a messaging protocol, not a blockchain. The protocol is designed to enable cross-chain communication and transfers \cite{Zarick2021}. It has a dual security model with layers of oracles and relayers. Oracles ensure the data integrity and verify message block headers. Relayers transmit and execute messages across chains and verify transaction proofs. Oracles and relayers are independent entities in the off-chain domain. The validation step will succeed if and only if the block header and the transaction proof match. It provides a strong trustless property.

\begin{figure}[htbp]
\centering
\includegraphics[width=\linewidth]{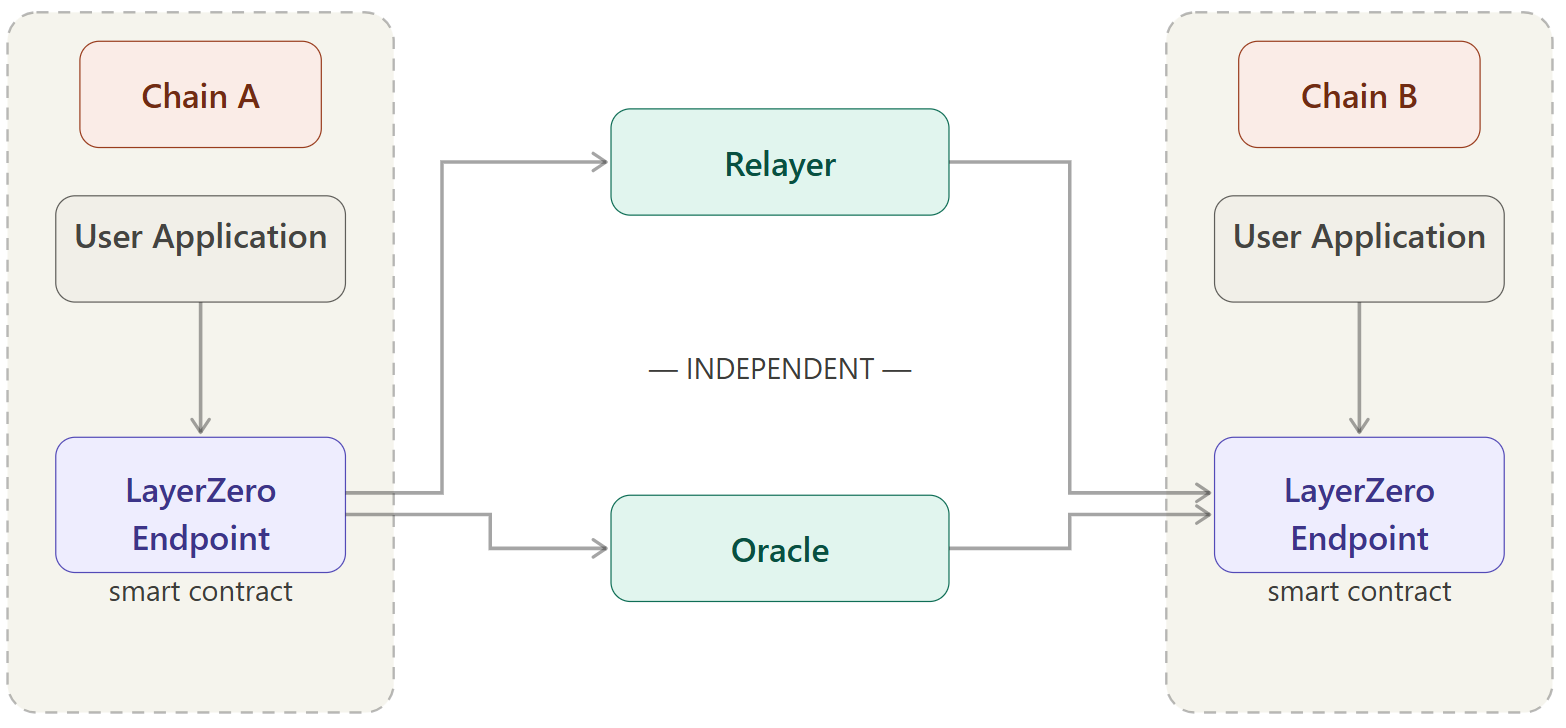}
\caption{LayerZero protocol is explained with the communication flow of a single transaction. Source: Zarich et al. (2021) \cite{Zarick2021}.}
\label{fig:layerzero_diagram}
\end{figure}

The first version of LayerZero integrated with Chainlink's 'decentralized oracle networks' (DONs) for enhanced security \cite{Zarick2021}. DONs operate under the Byzantine Fault Tolerance consensus mechanism. The second version of LayerZero introduces permissionless 'executors' that activate relayers and enhance permissionlessness. The sum of all these components is called 'omnichain'. The design is chain-agnostic. So it also provides interoperability between the Ethereum Virtual Machine (EVM) and non-EVM chains. Composed contract calls are isolated from each other. A manipulation by source contract functions doesn't affect the destination contract function. It means no cross-chain contamination.

LayerZero aims to be non-custodial and trust-minimized, which aligns with permissionless principles. However, the degree to which it is permissionless can depend on the specific implementation and use case. Developers can send arbitrary data, external function calls, and tokens across chains via LayerZero omnichain while preserving their control over their applications. Thus, Layerzero's architecture is modular at the verification level while it is static at the transport layer.

\subsection{Chainlink's Cross-chain Interoperability Protocol (CCIP)}\label{subsec2}

The CCIP is an arbitrary messaging protocol \cite{CCIP}. The CCIP is designed to enable secure messaging and token movements across different blockchains. It relies on DONs, which operate under the BFT consensus mechanism, to ensure data integrity and verify message block headers. Chainlink's oracle network is permissionless. Another component of the CCIP is the 'risk management network'. It is a secondary validation service parallel to the primary CCIP system. It inspects and scans the issues. A risk management node independently reconstructs the Merkle tree by fetching all messages on the source chain. Then, it checks for a match between the Merkle root committed by the "committing DON" and the root of the reconstructed Merkle tree.

\begin{figure}[htbp]
\centering
\includegraphics[width=0.8\linewidth]{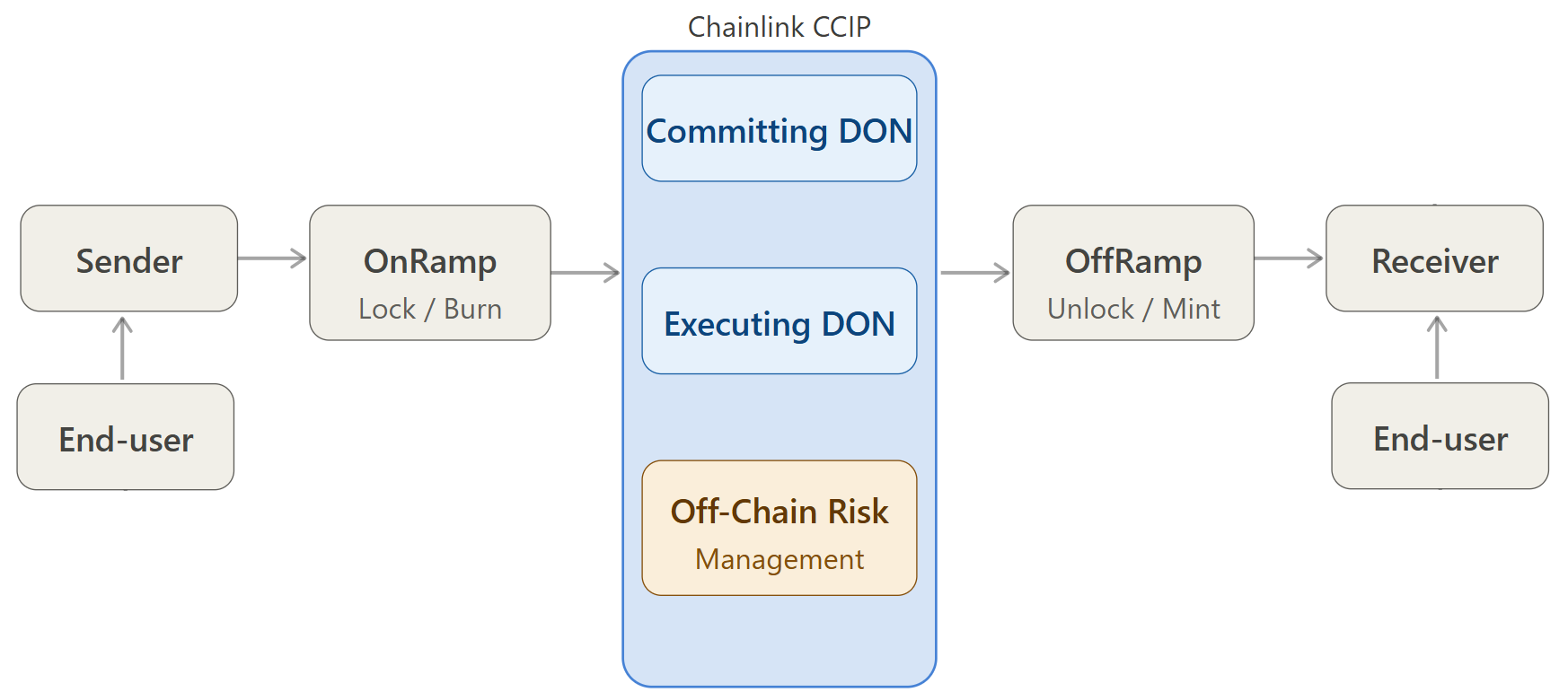}
\caption{The architecture of the Chainlink's CCIP. Source: Chainlink \cite{CCIP}.}
\label{fig:chainlink3}
\end{figure}

Any blockchain can connect and use the related cross-chain services. Chainlink targets the integration and API services for different type of private DLTs for further implications. But the CCIP is available for only EVM-compatible DLTs, as of 1st June 2024. Chainlink CCIP supports three main capabilities:

\begin{itemize}
    \item Arbitrary messaging is the ability to send arbitrary data (encoded as bytes) to a smart contract on a different blockchain. The developer can modify and encode any data. Usually, developers use arbitrary messaging to trigger an action on the receiving smart contract, including rebalancing an index, minting a specific non-fungible token, or calling an arbitrary function with the sent data as custom parameters. Developers can encode multiple instructions in a single message, enabling them to orchestrate complex, multi-step, multi-chain tasks.
    \item Token Transfer: Tokens can be transferred to a smart contract or directly to an externally owned account on a different blockchain. The CCIP uses burn-mint and luck-unlock mechanisms for cross-chain token transfer. 
    \item Programmable token transfer is the ability to simultaneously transfer tokens and arbitrary data within a single transaction. This mechanism allows users to transfer tokens and send instructions on what to do with those tokens. For example, a user could transfer tokens to a lending protocol with instructions to leverage those tokens as collateral for a loan, borrowing another asset to be sent back to the user.
\end{itemize}

\subsection{Circle's Cross-Chain Transfer Protocol (CCTP)}\label{subsec3}

The CCTP is a trusted messaging protocol natively for the USDC token \cite{CCTP}. USDC is an electronic money token issued by Circle and the CCTP is introduced to facilitate the secure and efficient transfer of USDC across various blockchain networks \cite{USDC}. It has a gateway API model. It employs a burn-and-mint mechanism for USDC transfers within the on-chain finance ecosystem.

While some networks have built-in protocols to transmit data across their constituent blockchains (e.g. Cosmos uses the Inter-Blockchain Communication (IBC) protocol to send information between its appchains), it is not possible for isolated networks, such as Ethereum and Avalanche, to communicate directly. The CCTP plays a role here. How it works:

\begin{itemize}
    \item USDC is burned on the source chain: A user initiates a transfer of USDC from one blockchain to another, by using an application. Different applications exist for non-EVM domains. The user specifies the recipient wallet address on the destination chain. The application facilitates a burn of the specified amount of USDC on the source chain.
    \item Circle observes and attests to the burn event on the source chain. The application requests the attestation from Circle, which provides authorization to mint the specified amount of USDC on the destination chain. This attestation service is used via a Gateway API that is directly controlled by Circle. The gateway is also integrated with other interoperability solutions like Chainlink’s CCIP.
    \item The application uses the attestation to trigger the minting of USDC. The specified amount of USDC is minted on the destination chain and sent to the recipient wallet address. 
\end{itemize}

Developers can build cross-chain applications that stack together the various functionalities of trading, lending, payments, non-fungible tokens (NFTs), and gaming. Users can perform cross-chain swaps with digital assets that live on disparate chains in an automated way. Users don't need to switch wallets or even pay attention to which chain the USDC is held. It shields the user from complexity.

\subsection{Wormhole}\label{subsec4}

Wormhole is a cross-chain messaging protocol that allows for the transfer of value and information between different blockchains \cite{Wormhole}. It is designed to be considerably more decentralized and permissionless. It has several components in its architecture. A decentralized validator network, which is called the 'guardian network', observes messages and signs the corresponding payloads. Each node performs in isolation. This network acts according to the core contract of the protocol. The off-chain process resumes through two relayer networks. One relayer network delivers messages of requests. Another relayer network executes them.

Cross-chain token transfer functions with burn-mint and lock-unlocked mechanisms. Wormhole is chain-agnostic. It has a modular architecture, it is not only EVM compatible.

\subsection{Connext}\label{subsec5}

Connext is a modular cross-chain messaging protocol \cite{Connext}. It has several components in its architecture. Connectors are on-chain smart contracts that are deployed on spoke domains to implement data transfer methods to send the Merkle root of all messages that originate from the spoke domain to the hub domain. Connectors are also deployed to send the aggregated Merkle root of all the received spoke Merkle roots to the configured destination domain.

The architecture of Connext has routers, sequencer, and relayers for the off-chain process. Routers are liquidity providers to enable instant liquidity for cross-chain token transfer on the destination chain in return for a fee. It has a permissionless nature, any user can become a liquidity provider. A sequencer collects bids from the integrated chains and randomly selects routers to fulfill the bids. Then the sequencer posts batches of these bids to a relayer network to submit them to the destination chain. Finally, a relayer network executes smart contracts in the chain. Connext is using Gelato as a relayer service, as of 1st June 2024 \cite{Gelato}.

For instance, in an EVM-compatible cross-chain data transfer, a message passed between Polygon and Optimism is secured by a proof that is posted to Ethereum and verified by the Polygon PoS bridge and Optimism roll-up bridge. This mechanism gives developers better possible trust guarantees for whichever chains they want to build on. If fraud occurs, the system falls back to using the canonical messaging bridge for each chain ecosystem. But Connext is chain-agnostic, not only EVM compatible. Similarly, a message passed within the Cosmos ecosystem is verified by IBC. More than one verification cluster exists in Connext. Consequently, application developers can choose different verification mechanisms for different chains.

\subsection{Hop Protocol}\label{subsec6}

Hop Protocol enables scalable and efficient rollup-to-rollup token transfers through a dual approach involving specialized bridge tokens and Automated Market Makers (AMMs) \cite{Whinfrey2021}. Ethereum rollups -they are separate blockchains in fact but orchestrated by a native smart contract on Ethereum- create siloed environments for its applications. Moving assets between rollups and the layer-1 network is slow and expensive. The Hop protocol allows assets to be moved directly from rollup to rollup by diminishing cost savings and enabling cross-rollup composability of applications.

\begin{figure}[htbp]
\centering
\includegraphics[width=0.8\linewidth]{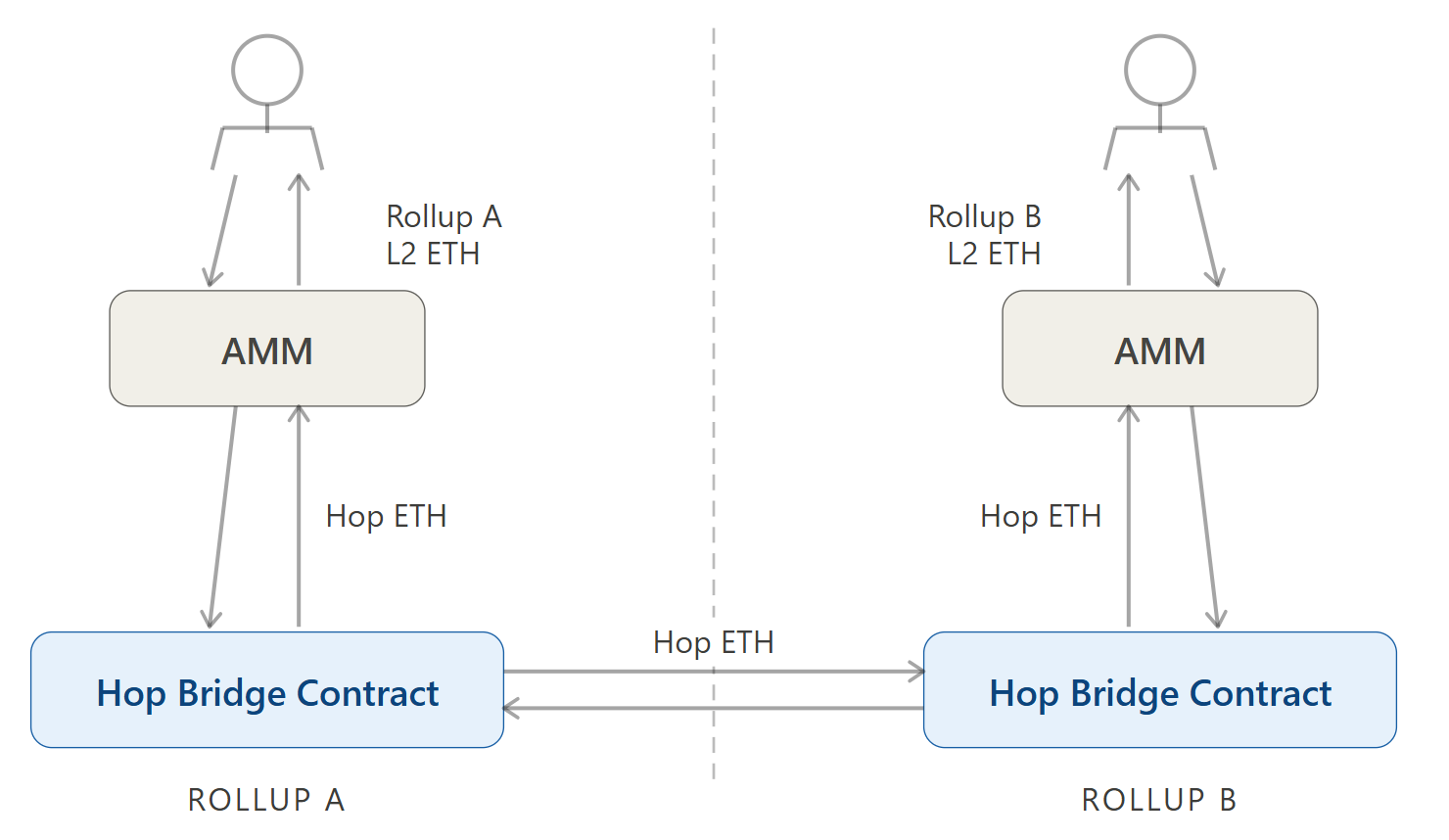}
\caption{Illustration of canonical token transfer by using the Hop token as an intermediary asset. Source: Whinfrey (2021) \cite{Whinfrey2021}.}
\label{fig:hop4}
\end{figure}

The protocol introduces Hop Bridge Tokens, such as Hop ETH and Hop DAI, which are minted on layer-2 when assets are deposited into a Layer-1 Hop Bridge contract and are burned when redeemed on layer-1. For rollup-to-rollup transfers, these tokens are burned on the origin rollup and minted on the destination rollup. Bonders provide immediate liquidity on the destination rollup in exchange for a small fee, which is later restored when the transfer is confirmed on layer-1. Transfers are aggregated into transfer roots, which bundle thousands of transfers into a single layer-1 transaction, enhancing scalability. On each rollup, AMMs facilitate the swapping between Hop Bridge Tokens and their corresponding canonical tokens, with pricing mechanisms and arbitrage opportunities ensuring liquidity rebalancing. Liquidity providers contribute to AMM pools and earn fees with minimal risk of impermanent loss due to the narrow price range of paired assets. This architecture allows users to efficiently perform rollup-to-rollup transfers with minimal layer-1 interactions and offers convenience functions for single-transaction cross-rollup swaps. As it is seen, the protocol can transfer tokens only between EVM rollups.

\subsection{Across}\label{subsec7}

Across is a cross-chain messaging protocol supported by a network of guardians and relayers for validating and relaying transactions, which enhances security and decentralization \cite{Across}. The guardian network ensures the validity and security of transactions, while relayers focus on efficient and accurate execution across different blockchain networks. Verification of off-chain data is reliant on an optimistic oracle network UMA \cite{UMA}. The protocol supports high transaction throughput and low latency, often integrating with layer-2 solutions like rollups to reduce gas fees and enhance scalability. It allows for multi-chain asset transfers and cross-chain messaging between EVM chains like Ethereum rollups.

\begin{figure}[htbp]
\centering
\includegraphics[width=0.9\linewidth]{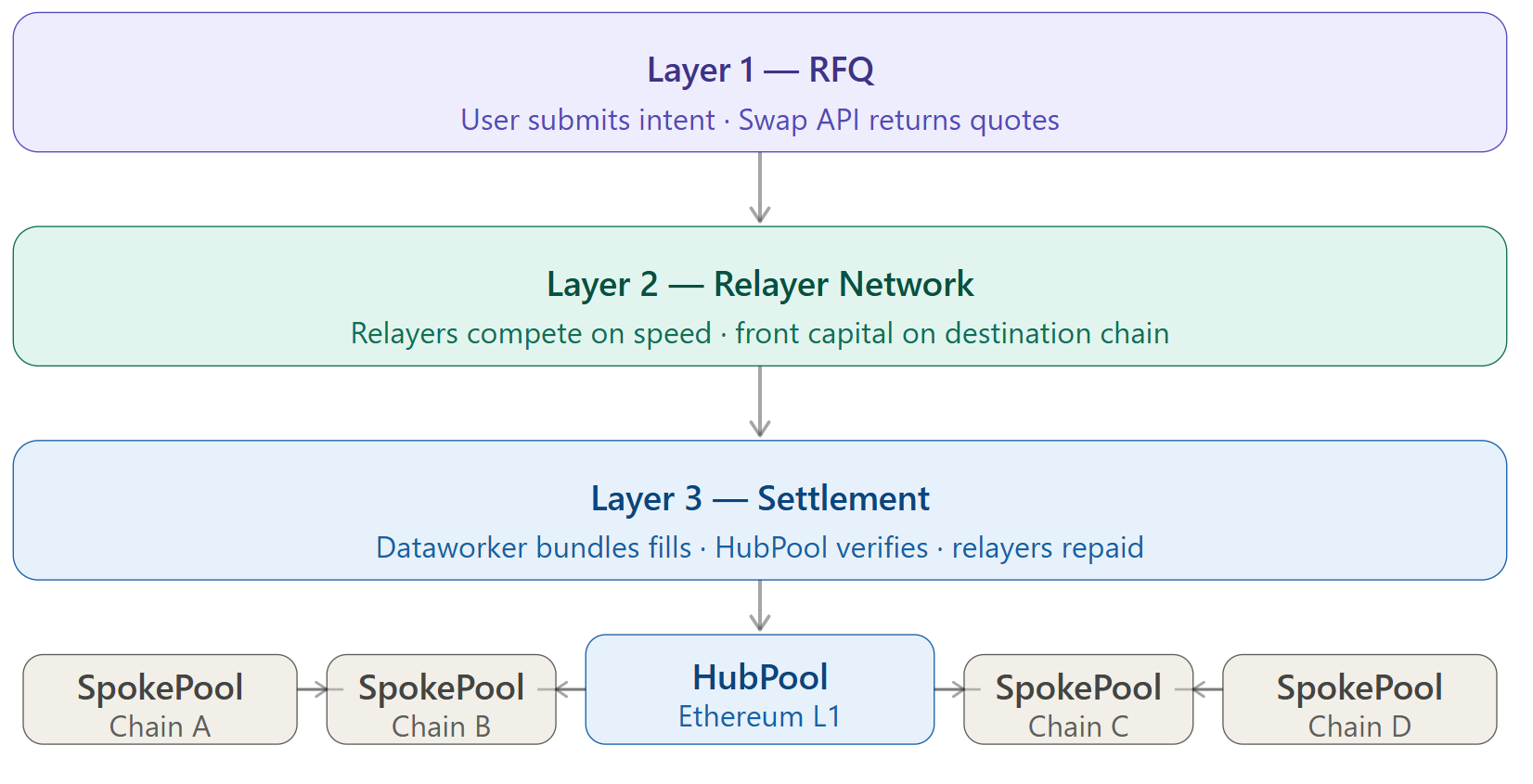}
\caption{A diagram of Across architecture. Source: \cite{Across}.}
\label{fig:across2}
\end{figure}

The technical architecture of Across includes smart contracts that govern the core functionality and handle the locking and releasing of assets during cross-chain transfers. It features a distributed validator network to ensure accurate and secure transaction execution. Bridges and relays facilitate communication between different blockchains, maintaining security and efficiency. The protocol also includes mechanisms for monitoring and verifying transactions to prevent fraud and ensure integrity.

Governed by a decentralized model, Across incentivizes participation through rewards for guardians and relayers, who play critical roles in validating and relaying transactions. This decentralized governance allows token holders to vote on protocol upgrades and changes, ensuring user involvement in its development. Across is particularly useful for DeFi applications, enabling users to move assets between different platforms, and for building cross-chain decentralized apps, providing developers with the tools to create applications that operate across multiple blockchains. Additionally, it supports asset portability, allowing users to transfer fungible and non-fungible tokens.

\subsection{Polkadot}\label{subsec8}

Polkadot is an early example of an open cross-chain ecosystem. In Polkadot, many domain-specific, parallel chains connect via a common relay chain that enables tokens and data to flow between them \cite{Wood2020}. Polkadot facilitates cross-chain interoperability through its relay chain and parachain architecture. The relay chain (Polkadot) uses nominated proof-of-stake (NPoS) consensus mechanism. Inter-chain communication always crosses this relay chain, thus incurring additional costs. 

\begin{figure}[htbp]
\centering
\includegraphics[width=0.8\linewidth]{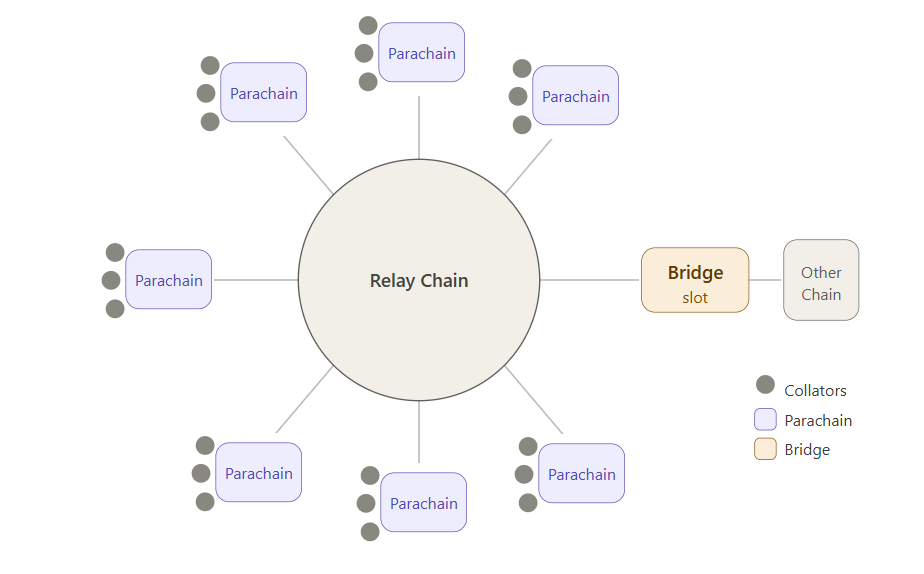}
\caption{Polkadot architecture. Source: \cite{Wood2020}.}
\label{fig:polkadot4}
\end{figure}

Parachains can be either permissioned or permissionless, depending on their governance structure. The Polkadot network as a whole is designed to be permissionless, allowing for decentralized participation in the network's security and governance. It is allowed to transfer arbitrary data across blockchains. It is possible to build applications that get permissioned data from a private blockchain and use it on a public blockchain.

\subsection{Cosmos}\label{subsec9}

Cosmos is a blockchain network technology that allows the sending of arbitrary messages between supported chains \cite{Cosmos}. Cosmos is built around the concept of an 'internet of blockchains' where each independent blockchain (called zones) can communicate with others via the Cosmos Hub that uses the Inter-Blockchain Communication (IBC) protocol \cite{IBC}. IBC is built on Tendermint BFT to facilitate messaging between chains built on Cosmos Hub. IBC enables interoperability in a considerably more decentralized manner. These features of IBC, combined with its use of an intermediate chain to facilitate consensus, make it similar to Polkadot, rather than a general communication layer. The chains depend on relayers to communicate. Relayers are the 'physical' connection layer of IBC. Off-chain processes are responsible for relaying data between two chains running the IBC protocol by scanning the state of each chain, constructing appropriate datagrams, and executing them on the opposite chain as it is allowed by the protocol. IBC can be used to build a wide range of cross-chain applications that include token transfers, atomic swaps, multi-chain smart contracts (with or without mutually comprehensible virtual machines), cross-chain account control, and data or code sharding.

\section{Comparison}\label{sec6}

Comparing the interoperability solutions provided by LayerZero, Wormhole, Chainlink CCIP, Circle CCTP, Polkadot, and Cosmos involves examining their approaches to facilitating communication and asset transfer between diverse blockchain ecosystems. The comparison is summarized in Table \ref{tab:cci-solutions-transposed}. The choice among them would depend on the specific needs of the application, including factors like the desired level of decentralization, security, ease of integration, and the types of assets or data being transferred. By utilizing this comparison, an empirical study can focus on the technical differences of these interoperability protocols, such as comparing the performance of chain-agnostic and non-chain-agnostic protocols, or the analysis of the metrics of the stablecoin-specific interoperability protocols like Circle CCTP.

\begin{table}[h]
\caption{A comparative table of the technical properties of cross-chain interoperability solutions, as of 1st June 2024.}\label{tab:cci-solutions-transposed}
\centering
\begin{tabular}{|p{1.35cm}|p{1.25cm}|p{2.5cm}|p{1.4cm}|p{1.15cm}|p{1.1cm}|p{1.1cm}|p{2.5cm}|p{2.5cm}|}
\hline
 & \textbf{Type} & \textbf{Key Features} & \textbf{Security Model} & \textbf{Modular} & \textbf{Trust Minimized} & \textbf{Chain Agnostic} & \textbf{Limits} \\ 
\hline
\textbf{LayerZero V1} & Messaging Protocol & Unique endpoints facilitating chain communication. & Duality: Oracle, Relayers & Partially Yes & Yes & Yes & Requires specific dApp integration; security depended on oracles/relayers \\ 
\hline
\textbf{Chainlink CCIP} & Messaging Protocol & Established oracle network for cross-chain communication. & Oracle Network, Risk Management Network & Currently No & Yes & Currently No & Reliant on oracle integrity; involves oracle risks \\ 
\hline
\textbf{Circle CCTP} & Transfer Protocol & Simplifies USDC transfers, enhancing liquidity. & Trusted Third-party Gateway & No & No & Yes & Focused on USDC transfers; reliant on burn-and-mint security \\ 
\hline
\textbf{Wormhole} & Messaging Protocol & Decentralized attestation model for transaction verification. & Validator Network, Relayer Network & Partially Yes & Yes & Yes & Security of validator nodes and risk of collusion or compromise \\ 
\hline
\textbf{Connext} & Messaging Protocol & State channel network for scalable cross-chain communication. & Connectors, Routers, Sequencer, Relayer network & Yes & Yes & Yes & Focused on token transfer; limited liquidity of routers \\ 
\hline
\textbf{Hop Protocol} & Transfer Protocol & Scalable rollup-to-rollup token transfers using specialized bridge tokens and AMMs. & Transfer liquidity provision secured by Layer-1 verification. & Yes & Yes & No & Relies on transfer liquidity providers; potential for delayed layer-1 settlement. \\ 
\hline
\textbf{Across V2} & Messaging Protocol & Focused on token transfers between spoke pools of rollups. & Validator network, relayer executers, oracle verification & Yes & Yes & No & Off-chain data verification is reliant on an optimistic oracle network \\ 
\hline
\textbf{Cosmos} & Blockchain & Hub-and-spoke model with IBC for blockchain communication. & Complexity of network-wide governance & Yes & Yes & Yes & Complexity of network-wide governance \\ 
\hline
\textbf{Polkadot} & Blockchain & Shared security model across parachains. & Blockchain of blockchains with BFT consensus & Yes & Yes & Yes & Relay chain capacity; parachain complexity \\ 
\hline
\end{tabular}
\end{table}

\section{Usage Metrics}\label{sec7}

To show interest in these novel cross-chain interoperability solutions, useful statistics are visualized here from the websites of LayerZero and Wormhole.\footnotemark The used metrics are mainly financial -like total value locked (TVL)- and user engagement metrics -such as user and transaction count- with the consideration that these output metrics are practically more useful to observe, measure, and compare the efficiency and success of the protocols. For the future of the interoperability solutions, using these metrics in analysis is also useful for detecting the existence of technical limits and potential optimization points in terms of liquidity provision and user experience.

\footnotetext{LayerZero and Wormhole open data can be found in the following links: \href{https://layerzeroscan.com/analytics/advanced}{https://layerzeroscan.com/analytics/advanced} and \href{https://wormholescan.io/}{https://wormholescan.io/} (last access on 1st June 2024).}

\begin{figure}[!htb]
    \centering
    \begin{minipage}{.49\textwidth}
        \begin{tikzpicture}
        \begin{axis}[
            title={LayerZero User Applications},
            xlabel={Time},
            ylabel={},
            xmin=0, xmax=8,
            ymin=0, ymax=50,
            scaled y ticks=false, 
            xtick={0,1,2,3,4,5,6,7,8},
            xticklabels={July, ,Sept, ,Nov, ,Jan '24, ,Mar},
            ytick={0,10,20,30,40,50},
            xticklabel style={rotate=90},
            legend pos=north west,
            ymajorgrids=true,
            grid style=dashed,
        ]
        
        \addplot[
            color=black,
            mark=none,
            thick,
            smooth,
        ] coordinates {
            (0,1)(1,1)(2,3)(3,8)(4,15)(5,23)(6,33)(7,46)
        };
        \end{axis}
        \end{tikzpicture}
        \caption{The growth in LayerZero user numbers\\ from 1st July 2023 to 1st March 2024. Y‑axis \\ values are expressed in thousands for \\ readability. Reported values are approximate \\and intended for comparative interpretation \\ rather than exact measurement.\\ Source: LayerZero Scan \cite{LayerZeroScan}.}
        \label{fig:layerzero_user}
    \end{minipage}%
    \begin{minipage}{.49\textwidth}
        \begin{tikzpicture}
            \begin{axis}[
                title={LayerZero Transactions Over Time},
                xlabel={Date},
                ylabel={},
                xmin=0, xmax=12,
                ymin=0, ymax=500,
                xtick={0,1,2,3,4,5,6,7,8,9,10,11,12},
                xticklabels={Jan '23, Feb, Mar, Apr, May, Jun, Jul, Aug, Sep, Oct, Nov, Dec, Jan '24},
                ytick={0,100,200,300,400,500},
                yticklabel style={
                    /pgf/number format/fixed,
                    /pgf/number format/precision=5
                },
                scaled y ticks=false,
                xticklabel style={rotate=90},
                legend pos=north west,
                ymajorgrids=true,
                grid style=dashed,
            ]
        
            \addplot+[thick, no marks] plot coordinates {(0,10) (1,20) (2,30) (3,40) (4,50) (5,350) (6,200) (7,150) (8,100) (9,200) (10,250) (11,300) (12,350)};
            \addplot+[thick, no marks] plot coordinates {(0,8) (1,18) (2,28) (3,38) (4,48) (5,340) (6,190) (7,140) (8,90) (9,190) (10,240) (11,290) (12,340)};
            \addplot+[thick, no marks] plot coordinates {(0,6) (1,16) (2,26) (3,36) (4,46) (5,330) (6,180) (7,130) (8,80) (9,180) (10,230) (11,280) (12,330)};
            \addplot+[thick, no marks] plot coordinates {(0,4) (1,14) (2,24) (3,34) (4,44) (5,320) (6,170) (7,120) (8,70) (9,170) (10,220) (11,270) (12,320)};
            \addplot+[thick, no marks] plot coordinates {(0,2) (1,12) (2,22) (3,32) (4,42) (5,310) (6,160) (7,110) (8,60) (9,160) (10,210) (11,260) (12,310)};
        
            \legend{Aptos, Arbitrum, Avalanche, Base, Others}
            \end{axis}
        \end{tikzpicture}
        \caption{The transaction count on LayerZero \\ from January 2023 to January 2024. Y‑axis \\ values are expressed in thousands for \\ readability. Reported values are approximate \\and intended for comparative interpretation \\ rather than exact measurement.\\ Source: LayerZero Scan \cite{LayerZeroScan}.}
        \label{fig:layerzero_volume}
    \end{minipage}
\end{figure}

January 2023 was a pivotal date for LayerZero as the most required omnichain integrations of LayerZero were presented to the users. It is clearly seen in Figure \ref{fig:layerzero_user} and Figure \ref{fig:layerzero_volume} that the interoperability protocol between Ethereum rollups and other pioneering public blockchains has been used with rising interest. LayerZero has not presented an official product for zk-rollups and zero-knowledge-proof technologies, in early 2024, that have a serious impact on scalability recently. Thus, the upcoming integrations and increasing adoption of this omnichain solution in the industry are taken into account together with the rising trend on the charts; LayerZero and related technologies may have exponential growth statistically. These charts are based on data sourced from LayerZero's website. For a more robust and deep empirical analysis, detailed transaction data, user engagement metrics, and external market data would be required. 

\begin{table}[h!]
    \begin{minipage}[t]{0.47\linewidth}
        \centering
        \caption{Wormhole Cross-Chain Total Outflow Transaction Activity from 20 July 2023 to 8 March 2024. \\ Source: Wormhole \cite{Wormhole}.}
        \begin{tabular}{@{}lcr@{}}
            \toprule
            Source   & \% of Total Transactions & Transactions \\ 
            \midrule
            Solana   & 21.76\%                 & 522,416            \\
            Polygon  & 12.68\%                 & 304,495            \\
            BSC      & 10.45\%                 & 250,794            \\
            Arbitrum & 8.86\%                  & 212,798            \\
            Celo     & 7.11\%                  & 170,757            \\
            Aptos    & 5.65\%                  & 135,712            \\
            Ethereum & 5.29\%                  & 126,939            \\
            Avalanche & 5.28\%                 & 126,856            \\
            Sui      & 4.58\%                  & 109,955            \\
            Fantom   & 4.49\%                  & 107,862            \\
            \bottomrule
        \end{tabular}
        \label{fig:wormhole1}
    \end{minipage}
    \hspace{0.4cm}
    \begin{minipage}[t]{0.47\linewidth}
        \centering
        \caption{Wormhole Cross-Chain Transactions (Inflow from other chains to Ethereum) from 20 July 2023 to 8 March 2024. \\ Source: Wormhole \cite{Wormhole}.}
        \begin{tabular}{@{}lcr@{}}
            \toprule
            Target       & \% of Total Transactions & Transactions  \\ 
            \midrule
            Solana       & 65.85\%                  & 83,592        \\
            Sui          & 6.99\%                   & 8,877         \\
            Arbitrum     & 5.23\%                   & 6,643         \\
            BSC          & 4.75\%                   & 6,034         \\
            Sei          & 4.19\%                   & 5,315         \\
            Polygon      & 2.34\%                   & 2,973         \\
            Moonbeam     & 2.04\%                   & 2,589         \\
            Base         & 1.86\%                   & 2,355         \\
            Optimism     & 1.41\%                   & 1,794         \\
            Avalanche    & 0.96\%                   & 1,215         \\
            \bottomrule
        \end{tabular}
        \label{fig:wormhole2}
    \end{minipage}
\end{table}

\begin{table}[h]
    \begin{minipage}[t]{0.47\linewidth}
        \centering
        \caption{Wormhole Cross-Chain Inflow Volume Activity (From Ethereum to other chains) from 20 July 2023 to 8 March 2024. \\ Source: Wormhole \cite{Wormhole}.}
        \begin{tabular}{@{}lcr@{}}
        \toprule
        Target    & \% of Total Volume & Volume (\$)           \\ \midrule
        Sui       & 52.05\%            & \$1,160,771,342       \\
        Solana    & 29.89\%            & \$666,691,355         \\
        Arbitrum  & 4.32\%             & \$96,415,666          \\
        Moonbeam  & 3.56\%             & \$79,097,428          \\
        BSC       & 1.96\%             & \$43,632,772          \\
        Celo      & 1.95\%             & \$43,464,801          \\
        Aptos     & 1.73\%             & \$38,571,272          \\
        Sei       & 0.82\%             & \$18,295,081          \\
        Base      & 0.62\%             & \$13,777,613          \\
        Polygon   & 0.60\%             & \$13,324,085          \\
        \bottomrule
            \end{tabular}
            \label{fig:wormhole3}
    \end{minipage}
    \hspace{0.4cm}
    \begin{minipage}[t]{0.47\linewidth}
            \centering
            \caption{Wormhole Cross-Chain Total Outflow Volume Activity from 20 July 2023 to 8 March 2024. \\ Source: Wormhole \cite{Wormhole}.}
            \begin{tabular}{@{}lcl@{}}
            \toprule
            Source   & \% of Total Volume & Volume (\$)           \\ \midrule
            Ethereum & 37.20\%            & \$2,230,700,701       \\
            Solana   & 17.90\%            & \$1,072,000,777       \\
            Sui      & 17.15\%            & \$1,026,463,015       \\
            Arbitrum & 5.44\%             & \$326,304,908         \\
            BSC      & 3.92\%             & \$234,989,757         \\
            Moonbeam & 2.86\%             & \$171,224,870         \\
            Avalanche & 2.48\%            & \$148,896,924         \\
            Polygon  & 2.12\%             & \$127,236,585         \\
            Aptos    & 2.02\%             & \$121,036,864         \\
            Optimism & 1.91\%             & \$114,604,457         \\
            \bottomrule
            \end{tabular}
            \label{fig:wormhole4}
    \end{minipage}
\end{table}

Table \ref{fig:wormhole1}, \ref{fig:wormhole2}, \ref{fig:wormhole3}, and \ref{fig:wormhole4} demonstrate the high volume and usage of another pioneering interoperability solution Wormhole.

\begin{figure}[htbp]
    \centering
    \begin{subfigure}[b]{0.49\linewidth}
        \centering
        \includegraphics[width=\linewidth]{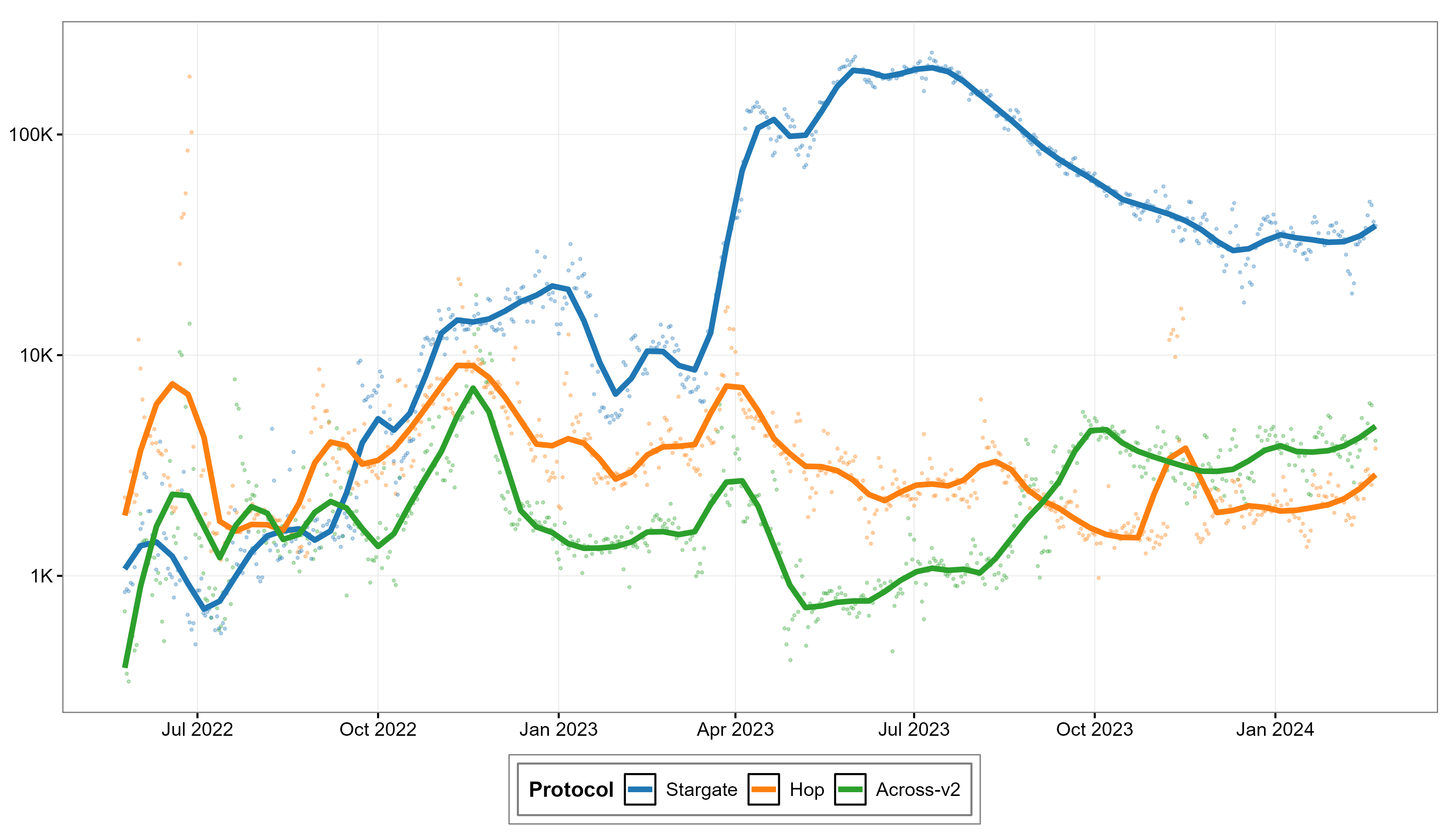}
        \caption{Daily active users by protocols.}
        \label{fig:daily_active_users}
    \end{subfigure}
    \hfill
    \begin{subfigure}[b]{0.49\linewidth}
        \centering
        \includegraphics[width=\linewidth]{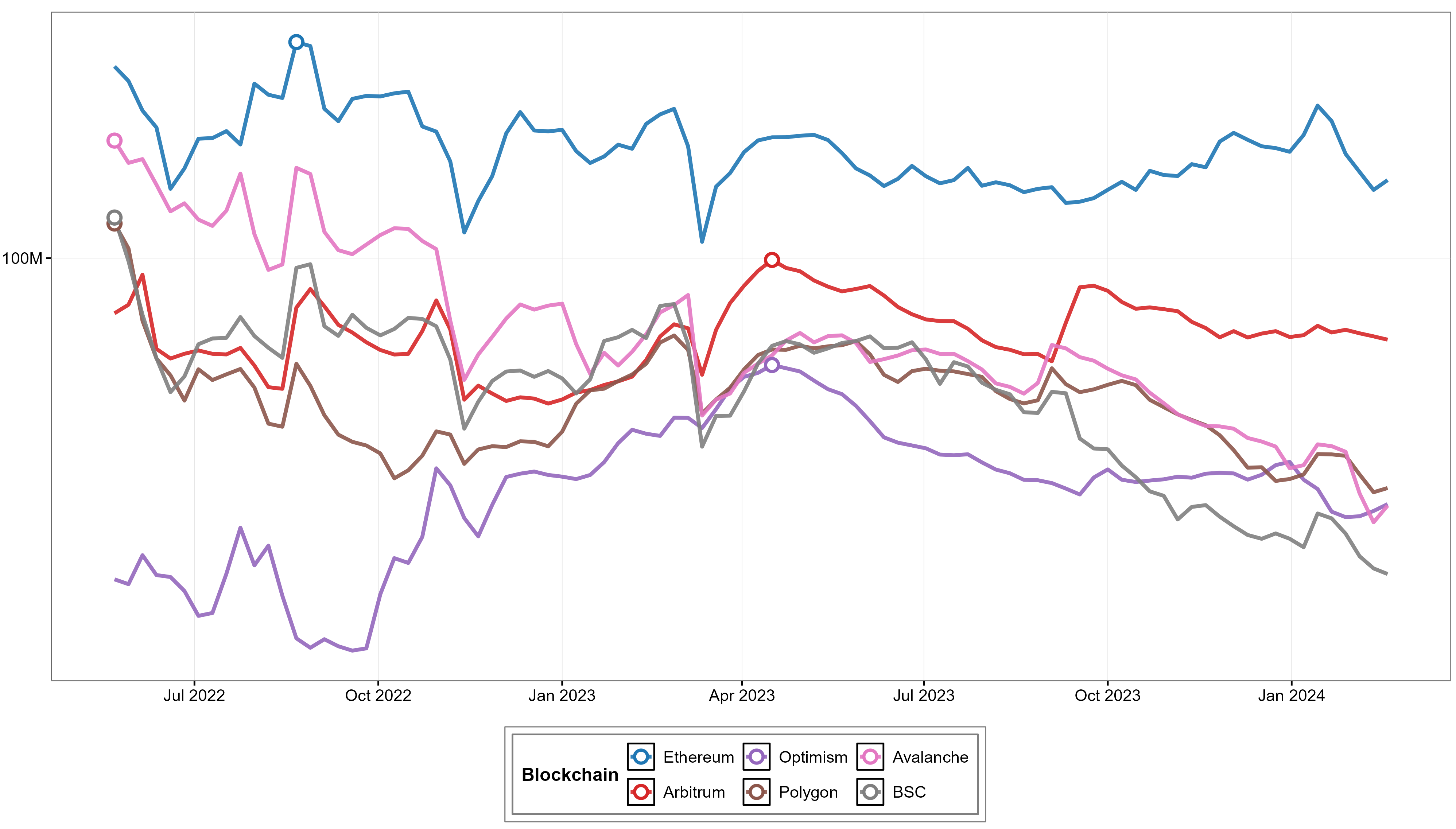}
        \caption{Total value locked by blockchains.}
        \label{fig:daily_tvl}
    \end{subfigure}
    \caption{The figure presents the comparison of daily active user count of bridge protocols (Stargate, Across v2 and Hop) and total value locked in the bridge protocols on different blockchains between 25th May 2022 and 21st March 2024. Data Source: Messari's Subgraphs on The Graph \cite{TheGraph}.}
    \label{fig:comparison_metrics1}
\end{figure}

Hop Protocol had been one of the most used cross-chain solutions as an early example of cross-rollup interoperability solutions, especially for Optimism rollup. However, as it's seen in Figure \ref{fig:daily_active_users}, the existence of competitors with chain-agnostic design and faster off-chain capabilities (such as LayerZero) would be the cause of the decline in the usage of Hop Protocol, specifically on the side of cross-chain liquidity providers. The contrary trend in the LayerZero charts above might be seen as proof of this argument.

These visualizations and tables can be a starting point for further research, like how to structure and analyze the context of DLT interoperability, how to model different scenarios or educational purposes, and case studies.

\begin{figure}[htbp]
    \centering
    \begin{subfigure}[b]{0.49\linewidth}
        \centering
        \includegraphics[width=\linewidth]{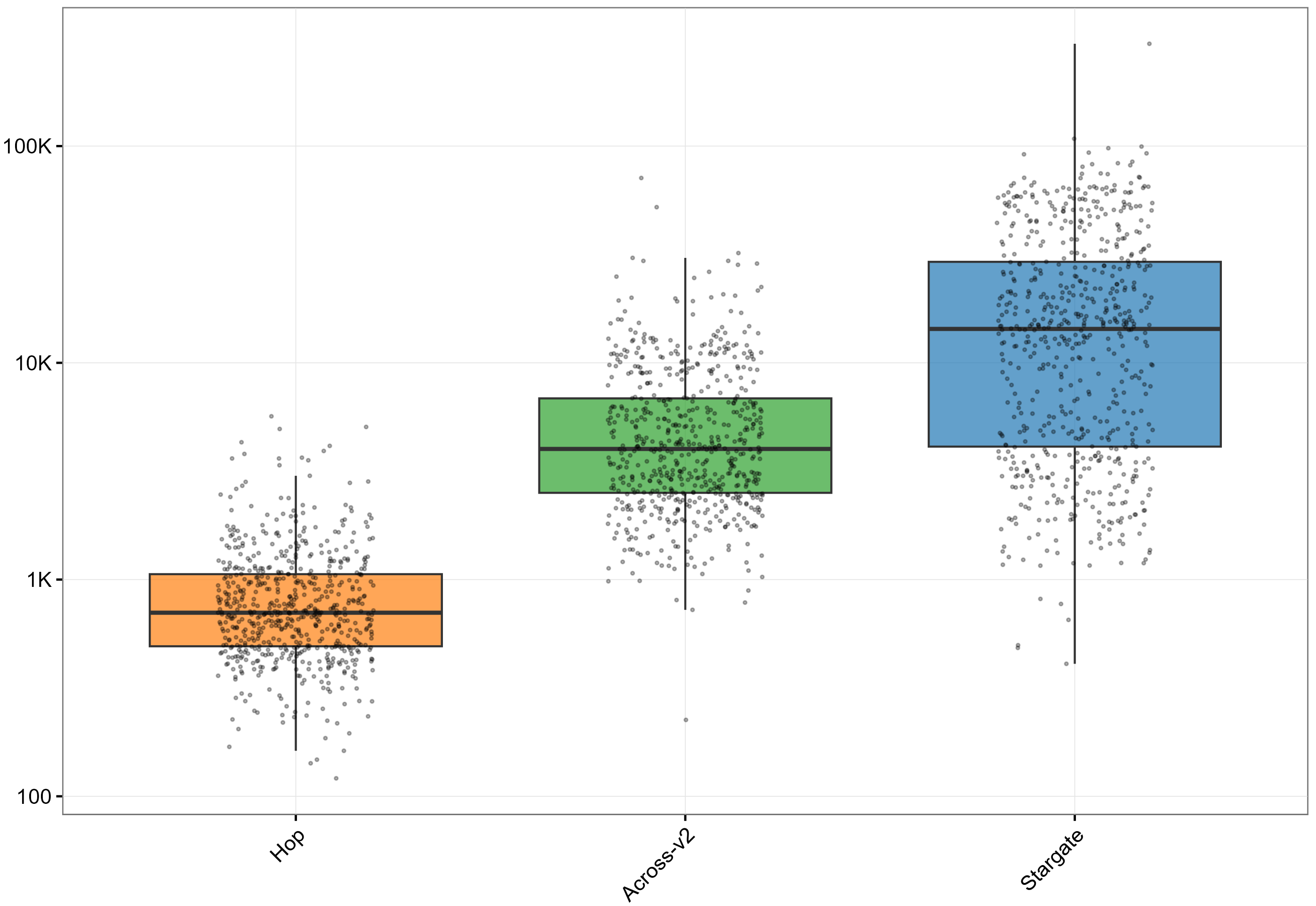}
        \caption{Daily revenue distribution by protocols (USD).}
        \label{fig:protocol_revenue_distribution}
    \end{subfigure}
    \hfill
    \begin{subfigure}[b]{0.49\linewidth}
        \centering
        \includegraphics[width=\linewidth]{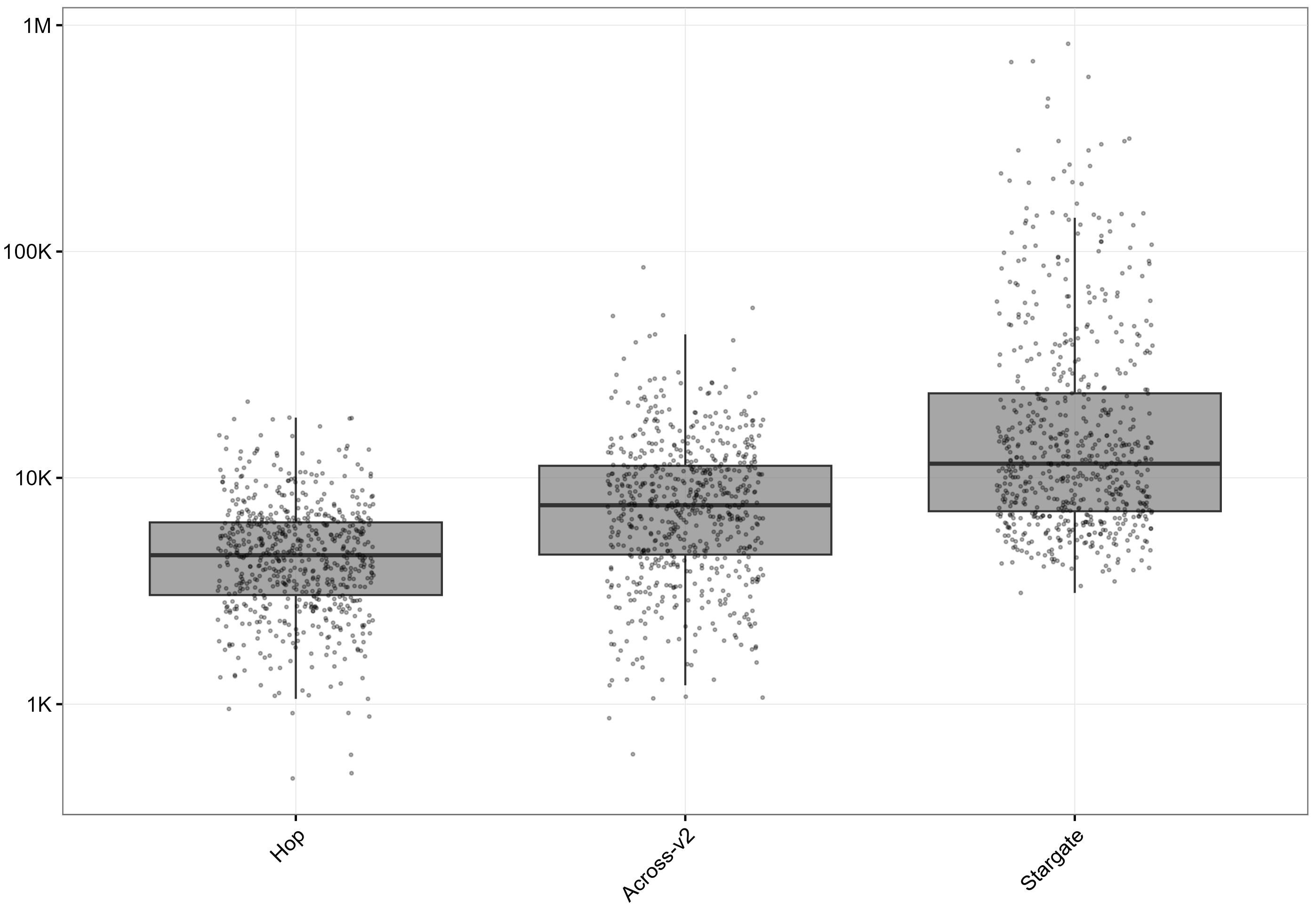}
        \caption{Daily volume distribution per tx by protocols.}
        \label{fig:protocol_volume_per_tx_distribution}
    \end{subfigure}
    
    \vskip\baselineskip
    
    \begin{subfigure}[b]{0.49\linewidth}
        \centering
        \includegraphics[width=\linewidth]{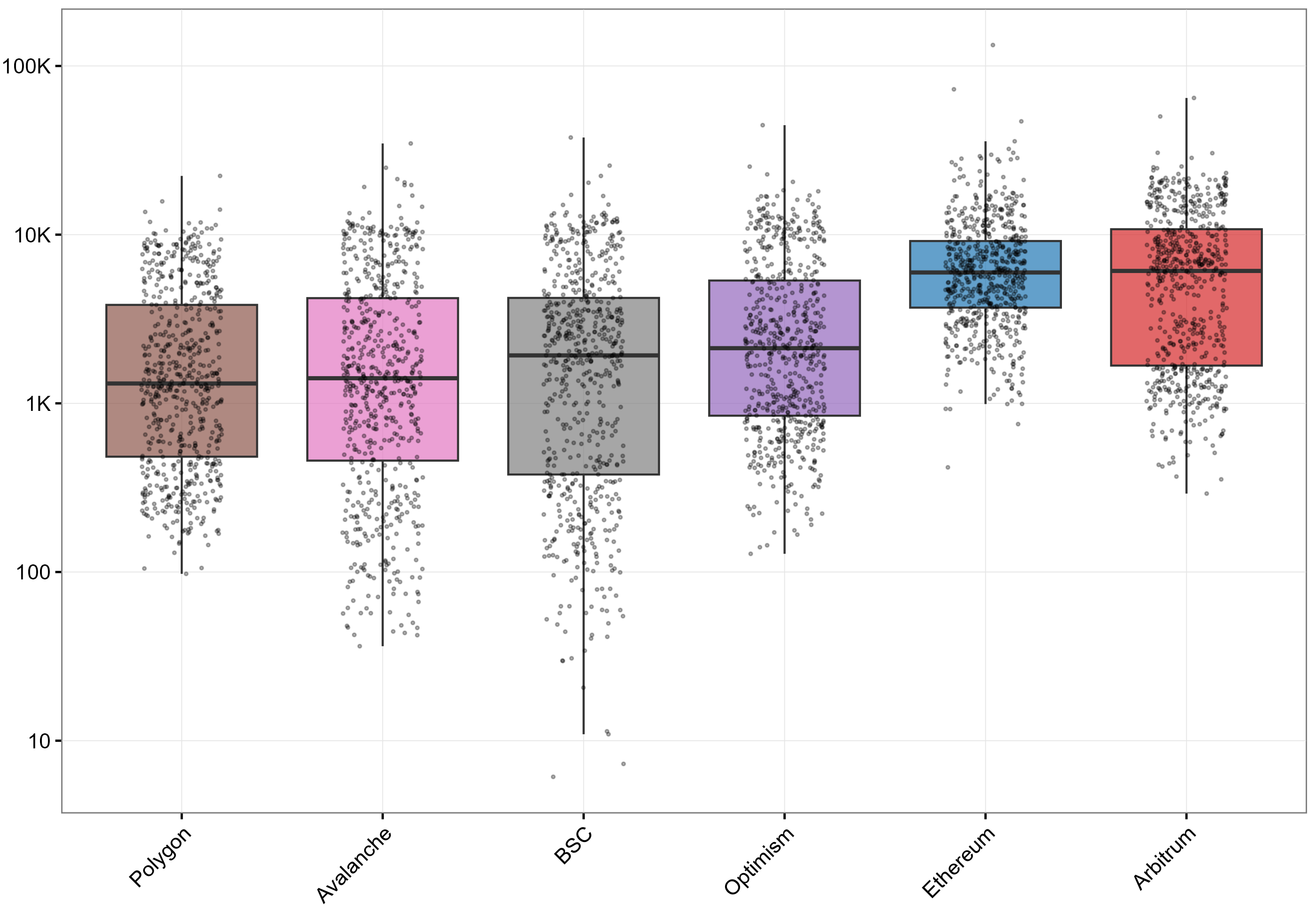}
        \caption{Daily revenue distribution by blockchains (USD).}
        \label{fig:chain_revenue_distribution}
    \end{subfigure}
    \hfill
    \begin{subfigure}[b]{0.49\linewidth}
        \centering
        \includegraphics[width=\linewidth]{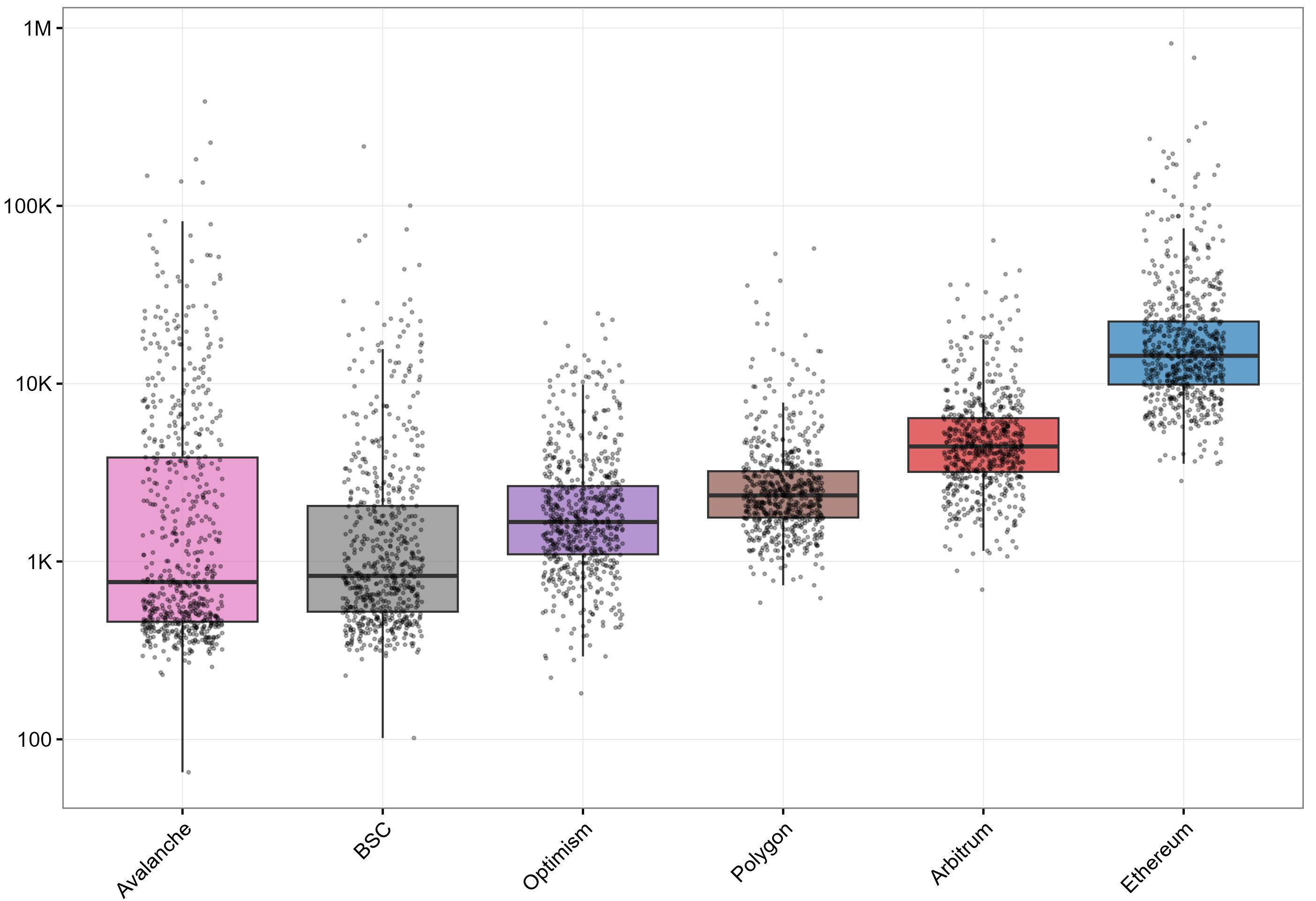}
        \caption{Daily volume distribution per tx by blockchains.}
        \label{fig:chain_volume_per_tx_distribution}
    \end{subfigure}
    \caption{The figures present box plots of different bridge protocols (Stargate, Across v2 and Hop) on different blockchains between 25th May 2022 and 21st March 2024. Data Source: Messari's Subgraphs on The Graph \cite{TheGraph}.}
    \label{fig:comparison_metrics2}
\end{figure}

\section{Empirical Comparison: Methodological Applications}\label{sec8}

The section attempts an example analysis with the data of cross-chain interoperability solutions: Across v2, Hop Protocol, and Stargate application \cite{Stargate}, which is the biggest application built on LayerZero's infrastructure. The data of three bridge protocols across 6 blockchains (Ethereum, Avalanche, Binance Smart Chain, Polygon, Arbitrum and Optimism) are collected for the period between 25th May 2022 and 21st February 2024 with daily time intervals.\footnotemark The selection of the protocols and blockchains is made according to the data availability of subgraphs on The Graph. The models and analyses are presented as an illustrative example rather than a comprehensive study. So the results should not be interpreted as definitive. The findings are based on data from only three interoperability platforms and a limited selection of variables. Expanding the scope to include more protocols and incorporating additional explanatory variables, such as fee structures, protocol incentives, or user retention metrics, could yield a more complete understanding of the dynamics within interoperability solutions.

\footnotetext{The data source is open and queried from \href{https://thegraph.com/explorer/subgraphs/7NAF7ZtNtJiXkfCFkTSAyFbfLLfUFa55UgK5woxPxZ46?v=0&view=Playground&chain=arbitrum-one}{Messari's Stargate Subgraphs}, \href{https://thegraph.com/explorer/subgraphs/KoMGmd2d1VMukusmf98SdWCSKH5ZP5UEUDHKrKpY74D?view=Overview&chain=arbitrum-one}{Messari's Hop Protocol Subgraphs}, and \href{https://thegraph.com/explorer/subgraphs/8c98cf7264e83e045d8b961a4e5ab3c84e08527b?view=Overview&chain=arbitrum-one}{Across v2 Protocol Subgraphs on The Graph}. A dataset is created by aggregating blockchain data from multiple deployments of each protocol across various blockchain networks. The rest of the subgraphs can be monitored in the following link: \url{https://thegraph.com/explorer/profile/0x7e8f317a45d67e27e095436d2e0d47171e7c769f?view=Subgraphs&chain=arbitrum-one} (last access on 1st June 2024).}

The models in Table \ref{tab:model_equations} are proposed to test the following objectives:

\begin{enumerate}
    \item To investigate whether aggregated revenue of bridge protocols and TVL in bridge contracts has a statistically significant relationship with net bridge volume and transaction activity, through baseline OLS regression models: Model 1 (M1) and Model 2 (M2).
    \item To evaluate whether the revenue elasticity to net flow volume differs significantly between Ethereum, layer-2s (Arbitrum, Optimism and Polygon), and alternative layer-1s (Avalanche and Binance Smart Chain), controlling for all time-invariant network characteristics and daily market shocks, through Model 3 (M3) fixed effects (FE) regression with panel data.
    \item To test the average relationship between revenue and its potential drivers (net volume, TVL, transaction count) for only Stargate protocol, masking potential protocol-specific heterogeneity, through Model 4 (M4) FE regression with panel data.
    \item To analyze how bridge protocols have different business models by estimating protocol-specific revenue elasticities to net flow volume on layer-2 networks, through Model 5 (M5) FE regression with panel data.
\end{enumerate}

\begin{table}[h!]
\centering
\caption{Model Specifications}
\label{tab:model_specs}
\begin{tabular}{p{3cm} p{11.5cm}}
\toprule
\textbf{Model} & \textbf{Specification} \\
\midrule
\textbf{M1:} Pooled OLS & $\text{TVL}_t = \beta_0 + \beta_1 \text{NetVolume}_t + \beta_2 \text{TxCount}_t + \beta_3 \text{GasPrice}_t + \beta_4 \text{ETH}_t + \epsilon_t$ \\[6pt]
\textbf{M2:} Pooled OLS & $\text{Revenue}_t = \beta_0 + \beta_1 \text{NetVolume}_t + \beta_2 \text{TxCount}_t + \beta_3 \text{TVL}_t + \beta_4 \text{GasPrice}_t + \beta_5 \text{ETH}_t + \epsilon_t$ \\[6pt]
\textbf{M3:} FE Chain-Type & $\text{Revenue}_{i,t} = \alpha_i + \lambda_t + \beta_1 \text{NetVolume}_{i,t} + \beta_2 [\text{NetVolume}\times \text{L2}]_{i,t}$ \\
& $+ \beta_3 [\text{NetVolume}\times \text{AltL1}]_{i,t} + \beta_4 \text{TxCount}_{i,t} + \beta_5 \text{TVL}_{i,t} + \epsilon_{i,t}$ \\[6pt]
\textbf{M4:} FE Network & $\text{Revenue}_{i,t} = \alpha_i + \lambda_t + \beta_1 \text{NetVolume}_{i,t} + \beta_2 \text{TxCount}_{i,t} + \beta_3 \text{TVL}_{i,t} + \epsilon_{i,t}$ \\[6pt]
\textbf{M5:} FE Protocol & $\text{Revenue}_{i,t} = \alpha_{p,c} + \lambda_t + \beta_1 \text{NetVolume}_{i,t} + \beta_2 [\text{NetVolume}\times \text{Across}]_{i,t}$ \\
& $+ \beta_3 [\text{NetVolume}\times \text{Hop}]_{i,t} + \beta_4 \text{TxCount}_{i,t} + \beta_5 \text{TVL}_{i,t} + \epsilon_{i,t}$ \\
\bottomrule
\label{tab:model_equations}
\end{tabular}
\end{table}

$Revenue$ represents the total revenue of a protocol. $TVL$ represents THE total value locked in protocols. $NetVolume$ is the sum of the total inflow and outflow volumes in the protocols. $TxCount$ refers to the total number of transactions conducted daily. $GasPrice$ represents the gas price of blockchain networks. $ETH$ represents the price of ETH as the benchmark asset in the on-chain finance. $L2$ and $AltL1$ indicate whether a variable belongs to layer-2 or alternative layer-1 networks. $AcrossV2$ and $Hop$ indicate whether a variable belongs to Across v2 or Hop Protocol bridge. All variables are used after logarithmic transformation. $NetVolume$ is log-like transformed with inverse hyperbolic sine (IHS) since it solves the non-positive data problem in log transformation \cite{Ravallion2017}. M1 and M2 are logarithmic OLS regression models while M3, M4 and M5 are logarithmic fixed effects models with panel data.  All models have low VIF values ($<$5), and there is no high mechanical correlation ($<$0.90).

Different metrics are visualized with plots over time in Figure \ref{fig:daily_active_users}, \ref{fig:daily_tvl}, \ref{fig:protocol_revenue_distribution}, \ref{fig:protocol_volume_per_tx_distribution}, \ref{fig:chain_revenue_distribution}, and \ref{fig:chain_volume_per_tx_distribution}. The summary statistics of the aggregated raw data are given in Table \ref{tab:summary_statistics_1}. The correlation matrix of the variables is presented in Table \ref{tab:correlation_matrix_1} while the VIF values are introduced in Table \ref{tab:VIF}. The hypothetical models in Table \ref{tab:model_equations} are tested. The regression results are presented in Table \ref{tab:regression_results_1}.

\begin{table}[h!]
\centering
\caption{The table presents summary statistics of aggregated data which involves three multi-blockchain bridge protocols and six blockchains. The number of observations is 638 without a missing value.}
\begin{tabular}{lcccccccc}
\hline
\textbf{Variable}             & \textbf{Mean}    & \textbf{Median}  & \textbf{SD}       &    \textbf{Min}       & \textbf{Max} & \textbf{Skewness}  & \textbf{Kurtosis} \\ \hline
Revenue (USD)             & 28.072         & 23.333           & 24.193             &    1.938       & 305.149       & 3.04               & 26.2              \\ 
Transaction Count               & 105.113        & 56.306           & 119.639            &    3.406       & 474.939       & 1.38               & 0.642              \\ 
Active Users                    & 59.205         & 37.084           & 60.912             &    2.321       & 239.426       & 1.26               & 0.400             \\ 
Total Volume (USD)              & 79.892.919     & 61.256.751       & 63.769.300          &    5.199.917    & 421.609.485    & 1.20               & 1.19              \\ 
Net Volume (USD)                & 478.107        & -99.173          & 6.120.211           & -30.064.826     & 69.653.516     & 2.38               & 27.7              \\ 
TVL (USD)        & 460.112.744    & 448.492.687      & 79.847.213          &    307.630.904  & 771.236.618    & 1.00               & 1.60              \\ \hline
\end{tabular}
\label{tab:summary_statistics_1}
\end{table}

\begin{table}[h!]
\centering
\caption{Variance inflation factors of the variables in the baseline model are represented.}
\begin{tabular}{lccccc}
\hline
\textbf{NetVolume} & \textbf{TxCount} & \textbf{TVL} & \textbf{GasPrice} & \textbf{ETH} \\
\hline
1.13 & 1.36 & 1.28 & 1.10 & 1.30 \\
\hline
\end{tabular}
\label{tab:VIF}
\end{table}

\begin{table}[h!]
\centering
\caption{Correlation matrix for the variables in the baseline model is represented.}
\begin{tabular}{|l|r|r|r|r|r|r|r|r|}
\hline
\textbf{Variable}  & Revenue & Tx Count  & Active Users  & Total Volume  & Net Volume    & TVL   & Gas Price & ETH Price \\ \hline
Revenue           & 1.00          & 0.73      & 0.73          & 0.91          & 0.02          & -0.28      & -0.04          & 0.41          \\ \hline
Tx Count                & 0.73          & 1.00      & 0.99          & 0.85          & -0.08          & -0.10      & -0.04          & 0.27          \\ \hline
Active Users            & 0.73          & 0.99      & 1.00          & 0.85          & -0.06          & -0.16      & -0.05          & 0.27          \\ \hline
Total Volume            & 0.91          & 0.85      & 0.85          & 1.00          & -0.01          & -0.19      & -0.04          & 0.33          \\ \hline
Net Volume              & 0.02          & -0.079      & -0.06          & -0.01          & 1.00          & -0.18      & 0.01          & 0.22          \\ \hline
TVL                     & -0.28          & -0.10      & -0.16          & -0.19         & -0.18          & 1.00  & 0.03          & -0.29          \\ \hline
Gas Price               & -0.04          & -0.04      & -0.05          & -0.04          & 0.01          & 0.03      & 1.00      & 0.03          \\ \hline
ETH Price               & 0.41          & 0.27      & 0.27          & 0.33          & 0.22          &  -0.29     & 0.03          & 1.00          \\ \hline
\end{tabular}
\label{tab:correlation_matrix_1}
\end{table}

\begin{table}[h!]
\centering
\caption{The table includes regression results for all models. Standard errors are in parentheses. Coefficients are rounded to appropriate precision. Stargate is assumed as the base protocol in M5. Ethereum is assumed as the base network in M3. Significance levels are denoted by: $^{***}$ p $<$ 0.001, $^{**}$ p $<$ 0.01, $^{*}$ p $<$ 0.05, $^{.}$ p $<$ 0.01.}
\label{tab:regression_results_1}
\begin{tabular}{l@{\hspace{0.6em}}c@{\hspace{0.4em}}c@{\hspace{0.4em}}c@{\hspace{0.4em}}c@{\hspace{0.4em}}c}
\toprule
 & \multicolumn{2}{c}{\textbf{Pooled OLS (Log-Log)}} & \multicolumn{3}{c}{\textbf{Fixed Effects (Log-Log)}} \\
\cmidrule(lr){2-3} \cmidrule(lr){4-6}
 & \textbf{M1: TVL} & \textbf{M2: Revenue} & \textbf{M3: Chain-Type} & \textbf{M4: Network} & \textbf{M5: Protocol} \\
\midrule
Intercept                 & 20.5018 (0.282)$^{***}$  & 5.5105 (3.136)$^{.}$    & -                         & -                         & -                         \\ 
\midrule
NetVolume                 & -0.0022 (0.001)$^{***}$  & -0.0003 (0.001)         & -0.0035 (0.001)$^{*}$     & -0.0194 (0.003)$^{**}$    & -0.0152 (0.003)$^{**}$  \\
\addlinespace
NetVolume:L2              & -                        & -                       & -0.0002 (0.003)           & -                         & -                         \\ 
\addlinespace
NetVolume:AltL1           & -                        & -                       & -0.0154 (0.001)$^{***}$   & -                         & -                         \\ 
\addlinespace
NetVolume:AcrossV2        & -                        & -                       & -                         & -                         & 0.0495 (0.004)$^{***}$   \\ 
\addlinespace
NetVolume:Hop             & -                        & -                       & -                         & -                         & 0.0216 (0.004)$^{***}$   \\ 
\addlinespace
TxCount                   & -0.0323 (0.006)$^{***}$  & 0.4847 (0.0139)$^{***}$ & 0.4763 (0.075)$^{**}$     & 0.4492 (0.102)$^{**}$     & 0.7578 (0.003)$^{***}$  \\ 
\addlinespace
TVL                       & -                        & -0.5079 (0.149)$^{***}$ & 0.4478 (0.304)            & 0.3819 (0.241)            & 0.1613 (0.020)$^{***}$  \\ 
\addlinespace
GasPrice                  & 0.0181 (0.007)$^{**}$    & 0.0222 (0.012)$^{.}$    & -                         & -                         & -                          \\ 
\addlinespace
ETH                       & -0.0907 (0.037)$^{*}$    & 1.1676 (0.096)$^{***}$  & -                         & -                         & -                          \\
\midrule
Adjusted R-squared        & 0.2141                   & 0.7381                  & 0.8228                    & 0.8036                    & 0.8598                    \\ 
\addlinespace
Within R-squared          & -                        & -                       & 0.0332                    & 0.2702                    & 0.6461                   \\ 
\addlinespace
Observations              & 638                      & 638                     & 3825                      & 3822                     & 5098                        \\
\bottomrule
\end{tabular}
\end{table}

The regression results in Table \ref{tab:regression_results_1} provide insights into the relationships between variables across blockchains and interoperability protocols. In Model 1, $NetVolume$ shows a statistically significant and negative relationship with $TVL$. This suggests that net volume may suggest a liquidity drawn corresponding to a reduction in value locked within the bridge protocols. In Model 2, $TxCount$ and $ETH$ have a statistically significant and positive relationship with $Revenue$, while $TVL$ shows a negative relationship. In Model 3, $NetVolume$ of $AltL1$ networks exhibits a statistically significant and negative relationship with $Revenue$, relative to the baseline on Ethereum. In Model 4, Stargate's $TxCount$ and $TVL$ demonstrate a statistically significant and positive relationship with $Revenue$. In Model 5, $AcrossV2$ and $Hop$ bridge protocols exhibit a statistically significant and positive relationship with $Revenue$, relative to the baseline in Stargate.

This suggests potential differences in user behavior and operational characteristics between protocols. These findings contribute to understanding how transaction activity, fund flows, and platform-specific factors interact in interoperability protocols.

The models and variables used here may oversimplify the complex mechanisms that govern these systems and some relationships may be misrepresented or incomplete. Additionally, potential limitations in data and methodology could affect the generalizability of the results. Robustness tests might be required.

The purpose of this study is to serve as an example framework for future empirical research in the field of cross-chain interoperability. By providing a foundational hypothetical model and tests, this analysis aims to inspire researchers to conduct more rigorous and comprehensive studies, leveraging extensive data sets and more customized methodologies to advance the understanding of interoperability solutions in decentralized ecosystems.

By comparing the metrics across platforms, these standard analyses are meaningful to evaluate which platform is performing better in terms of user engagement, transaction volume, and TVL. The plots in Figure \ref{fig:comparison_metrics1} and \ref{fig:comparison_metrics2} provide insights about the multi-blockchain protocols and landscape. Further econometric analyses, including spillover effects between different variables, regression models, Granger causality testing, and event studies, contribute to a comprehensive understanding of correlations and causalities in cross-chain finance and platforms.

\section{Conclusion and Future Research}\label{sec9}

This study contributes to the literature on distributed ledger technologies in financial use cases by exploring novel DLT interoperability solutions, their potential role in enhancing financial efficiency, and promoting empirical studies on the financial aspects of cross-chain solutions. Through a comprehensive analysis, the study demonstrated the mechanisms of these solutions to facilitate communication and transaction execution across disparate blockchain networks, eliminating the need for classical intermediaries and reducing counterparty risks. Different interoperability architectures exist with different security and performance trade-offs. The argument on the rising interest in the multi-blockchain bridges is supported by the provided statistics.

The paper exemplified an empirical analysis and evaluation with a regression analysis of data from three cross-chain finance platforms across multiple blockchains. By demonstrating how qualitative comparisons and empirical methods can be integrated, this study provides a practical framework that researchers can refine and adapt, paving the way for deeper exploration and improved evaluations of interoperability solutions. For future research on the financial impact of cross-chain interoperability solutions, the following areas are recommended to quantitatively and qualitatively analyze based on the insights from the document:

\begin{itemize}
    \item Analyzing and measuring the impact of cross-chain communication protocols on decentralized finance, protocol performance, financial efficiencies, transaction volumes, and financial dependencies of different DeFi services and networks.
    \item Contributing to the existing analyses and measures on the hack of cross-chain bridges and interoperability protocols.
    \item Examining the challenges and financial implications of executing smart contracts that span multiple blockchain environments, including the issues related to differing virtual machines and programming languages.
    \item Contributing to the governance models and audit mechanisms that can support secure and transparent cross-chain interactions, including the role of decentralized autonomous organizations and other forms of governance in managing cross-chain protocols.
    \item Studying the design of incentive schemes that encourage participation and maintain the security and efficiency of cross-chain networks, including the implications for transaction fees, token distributions in tokenization events, and staking rewards.
\end{itemize}

By addressing these areas, future research can contribute to the development of a mature, stable, and efficient cross-chain ecosystem, leading to greater value circulation, and innovative models for financial technologies.

\section{List Of Abbreviations}

DLT:     Distributed Ledger Technologies

CCC:     Cross-Chain Communication

CCTP:    Cross-Chain Transfer Protocol

CCIP:    Cross-Chain Interoperability Protocol

HTLC:    Hash-Time Locked Contracts

DEX:     Decentralized Exchange

USD:     United States Dollars

DON:     Decentralized Oracle Network

EVM:     Ethereum Virtual Machine

IBC:     Inter-Blockchain Communication

NFT:     Non-Fungible Token

AMM:     Automated Market Maker

NPoS:    Nominated Proof-of-Stake

OLS:     Ordinary Least Squares

TVL:     Total Value Locked

IHS:     Inverse Hyperbolic Sine

\end{document}